\documentclass[prd,aps,a4paper,floatfix,amsmath,amssymb,twocolumn,nofootinbib]{revtex4-1}
\usepackage{amssymb,amsmath,amsthm,xcolor,graphicx}
\usepackage[margin=2cm]{geometry}
\usepackage[pdftex,breaklinks,colorlinks,
linkcolor=blue,
citecolor=teal,
anchorcolor=red,
urlcolor=cyan]{hyperref}

\begin{document}


\title{Simulations of gravitational collapse in null coordinates:
  II. Critical collapse of an axisymmetric scalar field}


\author{Carsten Gundlach}
\affiliation{Mathematical Sciences, University of Southampton,
  Southampton SO17 1BJ, United Kingdom} 
\author{Thomas W. Baumgarte}
\affiliation{Department of Physics and Astronomy, Bowdoin College,
  Brunswick, ME 04011, USA} 
\author{David Hilditch}
\affiliation{CENTRA, Departamento de F\'isica, Instituto Superior
  T\'ecnico IST, Universidade de Lisboa UL, Avenida Rovisco Pais 1,
  1049 Lisboa, Portugal} 

\date{24 April 2024}


\begin{abstract}  
We present the first numerical simulations in null coordinates of  the collapse of nonspherical regular initial data to a black  hole. We restrict to twist-free axisymmetry, and re-investigate the  critical collapse of a non-spherical massless scalar field. We find  that the Choptuik solution governing scalar field critical collapse  in spherical symmetry persists when fine-tuning moderately  non-spherical initial data to the threshold of black hole  formation. The non-sphericity evolves as an almost-linear  perturbation until the end of the self-similar phase, and becomes  dominant only in the final collapse to a black hole. We compare with  numerical results of Choptuik et al, Baumgarte, and Marouda et al,  and conclude that they have been able to evolve somewhat more  non-spherical solutions. Future work with larger deviations from  spherical symmetry, and in particular vacuum collapse, will require  a different choice of radial coordinate that allows the null  generators to reconverge locally.
\end{abstract}

\maketitle
\tableofcontents


\section{Introduction}
\label{sec:scalarcollapseintro}


In a companion paper \cite{paper1} (from now on Paper~I), we have
reviewed various formulations of the Einstein equations in twist-free
axisymmetry on outgoing null cones emanating from a regular
centre, intending to simulate gravitational collapse with them. 

As a first physics application, here we re-investigate the problem of
twist-free axisymmetric scalar field critical collapse. By critical
collapse we understand the investigation of the threshold, along a
1-parameter family of initial data, between regular data that do and
do not form a black hole in their time evolution. In ``type-II''
critical collapse, increased fine-tuning to the threshold $p=p_*$ gives rise
to arbitrarily small black hole masses, arbitrarily large curvatures,
and in the limit, it is conjectured, to a naked singularity
\cite{GundlachLRR}.

We choose scalar field critical collapse because it provides a
continuous bridge from spherical symmetry, where simulations on null
cones work well, to the twist-free axisymmetric vacuum case that is
our long-term goal. But it is also of physical interest in its own
right.

In \cite{JMMGundlach1999} it was claimed, based on the numerical
solution of a mode ansatz, that all non-spherical linear perturbations
of the spherical scalar field critical solution decay, with the least
damped $l=2$ perturbation only decaying quite slowly. Subsequent
numerical time evolutions in axisymmetry in cylindrical coordinates by
Choptuik et al \cite{Choptuiketal2003} appeared to indicate a
slowly growing $l=2$ perturbation of slightly non-spherical critical
collapse that eventually leads to the formation of two centres of
collapse.

This tension between the results of \cite{JMMGundlach1999} and
\cite{Choptuiketal2003} was resolved by fresh simulations, in
axisymmetry in spherical coordinates by Baumgarte
\cite{Baumgarte2018}, which indicated that sufficiently small
non-spherical perturbations decay (in agreement with the mode
stability claimed in \cite{JMMGundlach1999}) but that larger
perturbations do grow (in agreement with \cite{Choptuiketal2003}).

Interestingly, both \cite{Choptuiketal2003} and \cite{Baumgarte2018}
observe that in a regime of small finite deviations from spherical
symmetry, the spherical Choptuik solution
\cite{Choptuik1993,MartinGundlach2003} is still observed as an
approximate critical solution but with a period that decreases from
$\Delta\simeq 3.44$ with increasing non-sphericity, and that the
critical exponent also decreases from $\gamma\simeq 0.374$.

The authors of \cite{Choptuiketal2003} and \cite{Baumgarte2018}, in
axisymmetry, achieved fine-tuning to $|p-p_*|\sim 10^{-15}$ and
$|p-p_*|\sim 10^{-13}$, respectively (except that \cite{Baumgarte2018}
achieved only $10^{-8}$ at the largest deviation from spherical
symmetry). Other investigations of non-spherical scalar field collapse
had achieved much less fine-tuning: \cite{Healeyetal2014} achieved
$\sim 10^{-3.5}$, and \cite{Deppetelal2019} $\sim 10^{-5.5}$, both for
non-axisymmetric initial datap, and \cite{CloughLim2016} achieved
$\sim 10^{-1}$ in axisymmetry. Incidentally,
\cite{Healeyetal2014,Deppetelal2019,CloughLim2016} also evolve initial
data that are non-smooth at $R=0$.

Reid and Choptuik \cite{Reid2023} have fine-tuned axisymmetric,
time-symmetric, antisymmetric in $z$ initial data to $\sim 10^{-14}$
and found that two (mirror-symmetric) centres of collapse form (as one
would expect). These start out in a highly non-spherical state
(although this was not quantified). However, critical phenomena were
observed with $\Delta$ and $\gamma$ as in spherical symmetry, and it
was concluded that there is no evidence of a growing non-spherical
mode (in the evolution of each separate centre of collapse).

Marouda et al \cite{Maroudaetal2024} have examined twist-free
axisymmetric collapse of a complex massless scalar field, fine-tuning
to $\sim 10^{-10}$. As they treat a slightly different matter model,
none of their families of initial data coincides exactly with those of
\cite{Choptuiketal2003} and \cite{Baumgarte2018}, but their results
are consistent with those of \cite{Choptuiketal2003} and
\cite{Baumgarte2018} in the sense that $\delta\Delta$ and
$\delta\gamma$ observed in their two non-spherical families fit into a
monotonic ordering of families by (inferred) non-sphericity. See
already Table~\ref{table:allcomparison} below.

The achievable level of fine-tuning matters, as the $l=2$ unstable
mode claimed in \cite{Choptuiketal2003} would be very slowly-growing:
its authors see no bifurcation for non-sphericity parameter
$\epsilon^2\le 2/3$ [see Eq.~(\ref{trydata}) below for a definition],
but they see one after 2 and 1.5 echos, respectively, for
$\epsilon^2=3/4$ and $5/6$, where 3 echos is the best that can be
achieved in double-precision arithmetic, with $|p-p_*|\sim 10^{-15}$.
\cite{Maroudaetal2024} see about 4 echos in their Family IV before
bifurcation. 

In principle, being in polar coordinates, our code, as presented in
detail in Paper~I, should resolve small deviations from spherical
symmetry as well as that of \cite{Baumgarte2018}, whereas our choice
of null gauge should be able to resolve the critical solution well,
even without the adaptive mesh refinement of
\cite{Choptuiketal2003}. It turns out that this is true.

The plan of the paper is as follows: In Sec.~\ref{section:setup} we
briefly summarise our metric ansatz, gauge choice and diagnostics,
give our two-parameter family (strength and non-sphericity) of initial
data, and describe the bisection to the threshold of collapse. In
Sec.~\ref{section:sphericalsymmetry} we describe in detail our results
for spherical critical collapse of a massless scalar field. In
particular, we plot the fine structure of the curvature and black hole
mass scaling laws with high accuracy and compare with the available
literature. 

Sec.~\ref{section:nonspherical} gives our results for non-spherical
(twist-free axisymmetric) critical collapse. Our initial data, being
set on an outgoing null cones, cannot be compared directly with those
of \cite{Choptuiketal2003,Baumgarte2018}. However, by comparing the
amplitude of the $l=2$ deviation of the scalar field $\psi$ from
spherical symmetry during the phase of the evolution where the
solution is well approximated by the Choptuik solution, we infer that
our most non-spherical family, $\epsilon_2=0.75$, compares to
$\epsilon^2=0.5$ of \cite{Choptuiketal2003,Baumgarte2018}. With the
gauge choice presented here we cannot fine-tune families of initial
data that are more non-spherical.

We conclude in Sec.~\ref{section:conclusions} with a discussion of the
physical results and an outlook for our code. 


\section{Setup}
\label{section:setup}


\subsection{Metric and field equations}


We state here the equations we solve numerically, with full details of
the formulation and its discretization given in Paper~I.
The field equations we want to solve are the Einstein equations
\begin{equation}
R_{ab}=8\pi\nabla_a\psi\nabla_b\psi,
\end{equation}
and the massless, minimally coupled wave equation
\begin{equation}
\nabla^a\nabla_a\psi=0.
\end{equation}
We use units where $c=G=1$.

We write the general twist-free axisymmetric metric in the form 
\begin{eqnarray}
\label{ymetric}
  ds^2&=&-2G\,du\,dx-H\,du^2 \nonumber \\
&&+R^2\left[e^{2S f}S^{-1}(dy+S\,b\,du)^2
+e^{-2S f}S\,d\varphi^2\right]. \nonumber \\ 
\end{eqnarray}
We assume that the central worldline $R=0$ is at $x=0$ and that
spacetime is regular there. We have defined $y:=-\cos\theta$, so that
the range $0\le\theta\le \pi$ corresponds to $-1\le y\le 1$. The
azimuthal angle $\varphi$ has range $0\le\varphi<2\pi$. The Killing
vector generating the axisymmetry is $\partial_\varphi$. (We use the
convention of equating vector fields with derivative operators.) In
(\ref{ymetric}) we have used the shorthand
\begin{equation}
S:=1-y^2=\sin^2\theta.
\end{equation}

Each surface ${\cal N}^+_u$ of constant $u$ is an outgoing null cone,
assumed to have a regular vertex.  Each surface ${\cal S}_{u,x}$ of
constant $u$ and $x$ is assumed to be spacelike, and has topology
$S^2$. The outgoing future-directed null vector field normal to ${\cal
  S}_{u,x}$ is $U:=G^{-1}\partial_x$, and is also tangent to the
affinely parameterised generators of ${\cal N}^+_u$.  The ingoing
future-directed null vector
normal to ${\cal S}_{u,x}$ is
\begin{equation}
\label{Xidef}
\Xi:=\partial_u -B\partial_x-Sb \partial_y,
\end{equation}
where we have defined the shorthand
\begin{equation}
\label{Bdef}
B:={H\over 2G}.
\end{equation}
As we see from (\ref{Xidef}) $B$ plays the role of a ``shift'' in the
$x$-direction, with $B$ and $b$ relating our time direction
$\partial_u$ to the ingoing null direction $\Xi$. $U$ and $\Xi$ are
normalised relative to each other as $\Xi^aU_a=-1$. See Fig.~1 of
Paper~I for a sketch of our coordinate null cones and these vector
fields.

As described in Paper~I, we can solve a subset of the Einstein
equations and the wave equation (which we call the hierarchy
equations) for $G$, $b$, $\Xi R$, $\Xi f$ and $\Xi\psi$ on one time
slice, given $R$, $f$ and $\psi$ there, and this can be done by
explicit integration in $x$ in the right order. $H$ and $\partial_u$
do not appear in the hierarchy equations except in the combination
$\Xi$, and $H$ remains undetermined.

This means that we have a time evolution scheme where $R$, $f$ and
$\psi$ are specified at $u=0$. We then find $G$, $b$, $\Xi R$, $\Xi f$
and $\Xi\psi$ , choose $B$ freely and thus obtain $R_{,u}$, $f_{,u}$
and $\psi_{,u}$, and then evolve $R$, $f$ and $\psi$ forward in $u$.

We discretize in $x$ and $u$ using finite differencing, and in $y$
using a pseudospectral expansion in Legendre polynomials (that is,
spherical harmonics restricted to axisymmetry).


\subsection{Gauge choice}


Our numerical domain is $0\le x\le x_\text{max}$, $0\le u\le
u_\text{max}$. We choose $H$ within a class of gauge choices with the
property that $R=0$ remains at $x=0$ (and is timelike and regular),
and that $x=x_\text{max}$ is future spacelike. This means that our
numerical domain is a subset of the domain of dependence of the
initial data on $0\le x\le x_\text{max}$, $u=0$, and no boundary
condition is required on the outer boundary $x=x_\text{max}$. We set
$H=G=1$ at $x=R=0$, so that the proper time along this ``central
wordline'' is given by $u$.

Furthermore, we impose that $x=x_0$ is marginally future spacelike, in
the sense that $H(u,x_0,y)\le 0$, with equality for some $y$, for some
fixed constant $x_0$ in the range $0<x_0\le x_\text{max}$. Our
expectation is that, if a part of the spacetime we are constructing
numerically is approximated by a self-similar critical spacetime (with
accumulation point on the central worldline $x=0$), then we can (by
trial and error) adjust $x_0$ so that $x=x_0$ remains close to the
past lightcone of the accumulation point of scale echos. The resulting
coordinate system will then zoom in on the accumulation point and
maintain resolution of the ever smaller spacetime features as it is
approached, without the need for adaptive mesh refinement.  This
approach to critical collapse in null coordinates was used in
\cite{Garfinkle1995,ymscalar}, using regridding, and in
\cite{Rinne2020,PortoGundlach2022} using a shift term, all in
spherical symmetry.

More specifically, for this paper we have settled on a class of gauges
we call ``local shifted Bondi gauge'' (from now on, lsB gauge),
defined by $R(u,x,y)=\bar R(u,x)$. Within that class, we choose the
``lsB4'' flavour, defined by the shift $B$ defined in (\ref{Bdef})
taking the value
\begin{equation}
\label{lsb-XiRmin}
B_\text{lsB4}=-{\Xi R-\min_y(\Xi R)\over R_{,x}}
+\left(1-{x\over x_0}\right)\left(-{\Xi R\over R_{,x}}\right)_{x=0}.
\end{equation}
(Note that in Paper I we extensively tested the slightly different
lsB2 gauge.) In this gauge, every surface of constant $x>x_0$ is
future spacelike. We refer the reader to Paper~I for more
details. Note that $B(u,x,y)$ is continous but has discontinuous
transverse derivative across the hypersurfaces $x=x_*(u)$ where the
$y$-location of $\min_y(\Xi R)$ jumps.


\subsection{Diagnostics}


For any closed spacelike 2-surface ${\cal S}$, we define its ``Hawking
compactness''
\begin{equation}
\label{CHaw}
C({\cal S}):=1+{1\over 2\pi}\int_{\cal S} \rho_+\rho_-\,dS,
\end{equation}
where $\rho_+$ and $\rho_-$ are the outgoing and ingoing null
divergence. It is related to the well-known Hawking mass by
\begin{equation}
\label{MHaw}
M({\cal S}):={1\over 2}\sqrt{A({\cal S})\over 4\pi}\,C({\cal S}),
\end{equation}
where $A({\cal S})$ is the area of ${\cal S}$.

The standard indicator of black hole formation used in numerical
relativity on a spacelike time slicing is the appearance of a
marginally outer-trapped surface (from now on, MOTS) embedded in a
time slice. An outer trapped surface is defined by $\rho_+\le 0$ at
every point (and hence $C>1$), and a MOTS by $\rho_+=0$ (and hence
$C=1$). As we have discussed in Paper~I, we cannot expect the ingoing
past light cone of a MOTS to converge to a point, and so we cannot
expect to find any MOTS embedded in a coordinate null
cone.

Instead we use the Hawking compactness $C(u,x)$ of our coordinate
2-surfaces ${\cal S}_{u,x}$ as an indicator of black-hole
formation. When $\max_xC(u,x)\ge 0.99$ is first reached during an
evolution, we use the Hawking mass $M$ of the ${\cal S}_{u,x}$ where
that happens as an indication of the ``initial mass'' of the black
hole. Similarly, we use $\max_xC(u,x)\le 0.01$ as an indicator of
dispersion. In spherical symmetry, this works perfectly well: $C=1$ is
actually equivalent to a MOTS, and this MOTS is approached uniformly
at all angles $(\theta,\varphi)$. Beyond spherical symmetry, for
current lack of a better alternative, we still use $C(u,x)$ to
distinguish between dispersion and collapse.

However, $C({\cal S})\ge 1$ is clearly only necessary, not sufficient,
for ${\cal S}$ to be outer-trapped, and we are not aware of any
rigorous results linking 2-surfaces with $C\simeq1 $ to black holes,
nor of their previous use in the numerical relativity literature
beyond spherical symmetry.

In lsB gauge, where $R=\bar R(u,x)$, the Hawking mass $M(u,x)$ of the
coordinate 2-surfaces ${\cal S}_{u,x}$ obeys $M_{,x}(u,x)\ge 0$. There
are separate integrals for $C(u,x)$ and $M(u,x)$, and this gives to
two separate ways of computing $C$, whose numerical results we denote
by $C$ and $\tilde C$, and similarly $M$ and $\tilde M$. Their
agreement is strong test of numerical accuracy, as they are
discretised in different ways, see Paper~I for details.


\subsection{Initial data}


The authors of \cite{Choptuiketal2003} investigate non-spherical
scalar field collapse in axisymmetry, using two families of initial
data on a Cauchy surface. One is time-symmetric, with the initial
value of the scalar field a Gaussian elongated along the rotation
axis, with equatorial symmetry. The other is approximately ingoing,
with $\psi$ antisymmetric under reflections through the equator. Like
the code of \cite{Baumgarte2018}, ours is optimised for a single
centre of collapse, while the antisymmetric data have two, so we
evolve only a family of data designed to be similar to the first
family of \cite{Choptuiketal2003}.

The time-symmetric initial data of \cite{Choptuiketal2003} and
\cite{Baumgarte2018} at $t=0$, written in our notation, are
\begin{equation}
\label{trydata}
\phi(0,r,y)=p\,e^{-{r^2(1-\epsilon_2y^2)\over d^2}}, \qquad
\phi_{,t}(0,r,y)=0.
\end{equation}
Here $r$ is the standard radial coordinate in a 3-metric assumed to
be conformally flat at $t=0$.
With $\epsilon_2>0$, the initial data are elongated along the symmetry
axis $y=\pm1$, and with $\epsilon_2<0$ they are squashed. In
\cite{Choptuiketal2003,Baumgarte2018}, $\epsilon_2$ is written as
$\epsilon^2$ and only positive values are considered, but there is no
reason not to consider $\epsilon_2<0$, hence our change of
notation. Given the initial data for the scalar field,
\cite{Choptuiketal2003,Baumgarte2018} obtain full initial data for the
Einstein-scalar system by making the initial 3-metric conformally flat
and solving the Hamiltonian constraint for the conformal factor.

As we set initial data on an outgoing null cone $u=0$ rather than a
time slice $t=0$, we cannot construct initial data giving exactly the
same solutions. Instead we identify the two data sets as different
cross-sections of an (approximate) solution of the flat spacetime wave
equation.  For this purpose, we consider the function
\begin{eqnarray}
\label{Phieps}
\phi(t,r,y)&:=&{1\over 2r}\Bigl[
(t+r)e^{-{(t+r)^2\over d^2}(1-\epsilon_2y^2)} \nonumber \\ 
&&-(t-r)e^{-{(t-r)^2\over d^2}(1-\epsilon_2y^2)}\Bigr].
\end{eqnarray}
For $\epsilon_2=0$ only, this a spherically symmetric analytic
solution of the scalar wave equation on flat spacetime. For any
$\epsilon_2$, it also reduces to (\ref{trydata}) at $t=0$. We now
identify the null cone $u=0$ where we impose initial data with the
conical cross-section $r=t-t_0-r=0$ of (\ref{Phieps}) for some
constant $t_0$, and we identify $r$ with $R$ on this
cross-section. Finally, we set an overall amplitude $p$. Hence we set
\begin{equation}
\label{psieps}
\psi(0,R,y)=p\,\phi(t_0+R,R,y),
\end{equation}
where $\phi$ is defined by (\ref{Phieps}).
We also set
\begin{equation}
f(0,x,y) = 0.
\end{equation}
This is simply for lack of an alternative likely to be closer to the
corresponding data at $t=0$ being conformally flat, and is at least
consistent with spacetime being flat in the limit $p\to 0$.
Finally, we set 
\begin{equation}
\label{Rinit}
R(0,x,y) = {x\over 2},
\end{equation}
which is essentially a gauge choice.

Formally, as $p\to 0$ and $\epsilon_2\to 0$, the solution
from these data corresponds to that arising from (\ref{trydata}), with
the identification $u=t-R$. We also initialise $R(0,x,y)=x/2$, which
can be considered our initial choice of an lsB gauge.

Obviously, with gravity and/or with $\epsilon_2\ne 0$, (\ref{Phieps})
is no longer an exact solution of the curved-spacetime wave equation,
and the solution evolved from the null initial data
(\ref{psieps}) no longer has a moment of exact time symmetry, but it
should approximate the solutions arising from (\ref{trydata}), with
$\epsilon_2$ playing a quantitatively similar role.

Without loss of generality we set $d=1$ to fix an overall scale. We
also choose $t_0=-5$, which means that the null initial data
(\ref{psieps}) represents a Gaussian that is well separated from the
centre and still essentially ingoing. In the flat-space, spherical
limit $\epsilon_2=0$, $p\to 0$ the wave reaches the centre at $u=-t_0$,
and the moment of time symmetry $t=0$ corresponds to $u+R=-t_0$. Hence
(always in this limit) our null initial data $u=0$ intersect $t=0$ at
$R=-t_0$. For $\epsilon_2=0$, we shall choose $x_\text{max}=11$, so that
$R=-t_0$ intersects our initial data surface.

We realised only near the completion of this paper, from the
convergence tests in Appendix~\ref{appendix:convergencetestsold}, that
for $\epsilon_2\ne 0$ and $t\ne 0$ (\ref{Phieps}) is not actually
single-valued at the origin $r=0$, but depends on $y$. Our null
initial data (\ref{psieps}) are therefore also not single-valued at the
origin. In Appendix~\ref{appendix:convergencetestsnew} we present an
analytic solution of the flat space wave equation valid for all values
of $\epsilon_2$ that also reduces to (\ref{trydata}) for all values of
$\epsilon_2$, and null initial data (\ref{analytictrydata}) derived
from this solution. These are the initial data we intended to
create. 

However, convergence tests in the strong-gravity regime of the
non-analytic null data (\ref{psieps}) with (\ref{Phieps}) show that
the breakdown of convergence is limited to a small region near the
origin, at early times. The code seems to smooth out the data to
something that then converges. The corrrected null data
(\ref{analytictrydata}) also converge, but preliminary tests show
that at $\epsilon_2=0.75$ they require very large values of
$x_\text{max}$ in near-critical evolutions. For both these reasons ---
the error in the non-analytic data does not seem to matter, and the
analytic data are much harder to evolve --- we present here only
evolutions with the non-analytic initial data, and we believe that the
following results are still reliable.


\subsection{Bisection to the threshold of black hole formation}


Beginning from a value of $p$ that disperses and one that collapses,
we find the threshold value $p_*$ by bisection. We use super$n$ as
shorthand for $\ln(p-p_*)\simeq-n$ and sub$n$ for
$\ln(p_*-p)\simeq-n$. (This makes sense because $p$ is dimensionless
and $p_*\sim 1$.) We can find $p_*$ to machine precision (15 digits),
and the accumulation point of echos $u_*$ to about 7 digits, but our
numerical values of $p_*$, $u_*$ and $x_*$ will agree with their
continuum value to fewer digits. This is a well-known feature of
numerical simulations of critical collapse. Below we give their values
at different resolutions as a rough indication of numerical error.

Beyond spherical symmetry, during bisection in $p_*$, or later when we
sample $\ln|p-p_*|$ more finely, we occasionally obtain inconclusive
evolutions. For small $\epsilon_2$ at least there is a heuristic
workaround: when the evolution stops as the time step goes below an
acceptable threshold, we always find that $\max_{x,y}\rho_->0$ as
well. (The expansion of ingoing null cones, which is negative in flat
spacetime, has become positive at at least one point, presumably
because the spacetime and our coordinate null cones have become highly
non-spherical.) We take this combination as a (completely heuristic)
indicator of collapse, even if $C=0.99$ has not yet been reached. At
the largest non-sphericity and highest resolution, this collapse
criterion also fails, and we replace it by $\max_x C(u,x)\ge 0.8$.

We cannot be sure that what we thus classify as collapse or dispersion
really is, but if we then see subcritical curvature scaling on the
``dispersion'' side, and approximate self-similarity on both sides, we
take this to be a strong indication that we really are bisecting to
the collapse threshold.  Furthermore, any mass-like quantity in
near-critical evolutions will also scale, in particular the Hawking
mass on the first surface ${\cal S}_{u,x}$ where $C=0.99$ (or
$C=0.8$), and seeing this scaling at least provides further
confirmation that we are bisecting to the black hole threshold, even
if the mass we measure is related to the black hole mass only by a
factor of order one.

Once we have an estimate for $p_*$ (which depends on all the numerical
parameters), we evolve super and subcritical values of $p$ that are
equally spaced in $\ln|p-p_*|$, with 30 values per power of 10, to
give us better-resolved plots of the mass and curvature scaling laws.


\section{Results in spherical symmetry}
\label{section:sphericalsymmetry}


We begin with the spherical case $\epsilon_2=0$, and run with $N_y=1$
in lsB4 gauge. By experimentation we find that $x_0=8.24$ allows
bisection down to machine precision, while keeping the first appearance
of $C\ge 0.99$ somewhere in the middle of the grid as the bisection
proceeds (in order to resolve the critical solution). In other words,
the past lightcone of the accumulation point of scaling echos is at
$x_*\simeq 8.24$. We also find that $x_\text{max}=11$ is required in
order to capture the location where $C$ first reaches the threshold
value of $0.99$ that we take as an indication of
collapse. We find $p_*\simeq 0.2402$.
 
We have checked that our code shows clean pointwise second-order
convergence with $\Delta x$ in the two evolutions forming our initial
bracket: the clearly subcritical $p=0.2$, and the clearly
supercritical $p=0.3$ (up to a little before our collapse criterion
stops the supercritical evolution).

As a basic check, we have evolved spherically symmetric initial data
also with $\bar N_y=3$. This makes no visible difference to the
scaling plots or critical solution (except where they dissolve into
round-off noise). However, round-off error in the spectral operations
triggers unphysical random nonspherical perturbations. These are then
smoothed out by the shrinking of the numerical grid, and then, once
their $x$-dependence is resolved, continue to evolve under the
continuum perturbation equations. The $l=2$ spherical harmonic
components of $\psi$ and $f$, which from now on we shall denote
$\psi_2$ and $f_2$, reach an amplitude $\sim 10^{-12}$ and then decay
in sub15, while they blow up in super15 during the final
collapse. This blowup may be explained by the physical instability of
gravitational collapse to non-spherical perturbations, even if these
perturbations were triggered here by numerical error.

We stop the bisection to $p_*$ after 50 iterations, where we are
essentially at machine precision. However, the scaling laws for the
initial black hole mass and maximum Ricci curvature become noisy well
before we reach machine precision, at about sub13 for the Ricci
scaling (see Fig.~\ref{fig:eps0_Tscaling}) and at about super12 for
the mass scaling (see Fig.~\ref{fig:eps0_Mscaling}).

In spherical symmetry and evolving with $N_y=1$, the source of the
randomness is presumably round-off error, becoming important in the
evolutions already three or so orders of magnitude before we reach
machine precision in $p$ itself.


\subsection{Fine structure of the mass and curvature scaling laws}


From Appendix~\ref{appendix:wiggles}, which clarifies the
  derivation given in \cite{Gundlach1997}, we expect the scaling laws
\begin{eqnarray}
\label{curvscaling}
-{1\over2}\ln T&\simeq&A+\gamma\ln(p_*-p)\nonumber
\\ && +f_T\left(\ln(p_*-p)+B\right), \\
\label{massscaling}
\ln M&\simeq&A+\gamma\ln(p-p_*)\nonumber \\ &&
+f_M\left(\ln(p-p_*)+B\right), 
\end{eqnarray}
for
\begin{equation}
T:=\max_\text{whole spacetime}|\nabla_a\psi\nabla^a\psi|,
\end{equation}
and the mass $M$ of the first MOTS detected on our time slicing. We
consider $T^{-1/2}$ because it has dimension length, like $M$, and we
use the natural logarithm for all our plots. Here $\Delta\simeq 3.44$
and $\gamma\simeq 0.374$ are universal constants, and $A$ and $B$ are
family-dependent constants. $f_T$ is a universal periodic function
with period $\Delta/(2\gamma)\simeq 4.60$ in $\ln(p_*-p)$. $f_M$ is
also periodic with the same period, and is universal with respect to
initial data in any time slicing (such as our null cone slicing) that
is compatible with discrete self-similarity (from now on,
DSS). However, $f_M$ does depend on the slicing, as well as on our
collapse criterion.

Note that the Ricci scalar is $|\text{Ric}|=8\pi T$, so
$\ln|\text{Ric}|$ and $\ln T$ differ only by a constant.

To look for these scaling laws, we plot $\ln M-(A+\gamma\ln|p-p_*|)$
against $\ln\ln|p-p_*|$, and similarly for $T$. We then expect to see
only the fine-structures $f_M$ and $f_T$. To uniquely fix $A$ for a
given family, we define $f_T$ to have zero mean. To fix $B$ modulo
periodicity, we let the minima of $-(1/2)\ln T$ coincide with the
minima of $\sin(\gamma\ln|p_*-p|)$. We find that for our family of
spherical initial data $A\simeq 1.113$, $B=0.592$, $\gamma=0.374$ and
$\Delta=3.44$ provide a good fit to (\ref{curvscaling}) for our family
of spherically symmetric initial data.

\begin{figure}
\includegraphics[scale=0.67, angle=0]{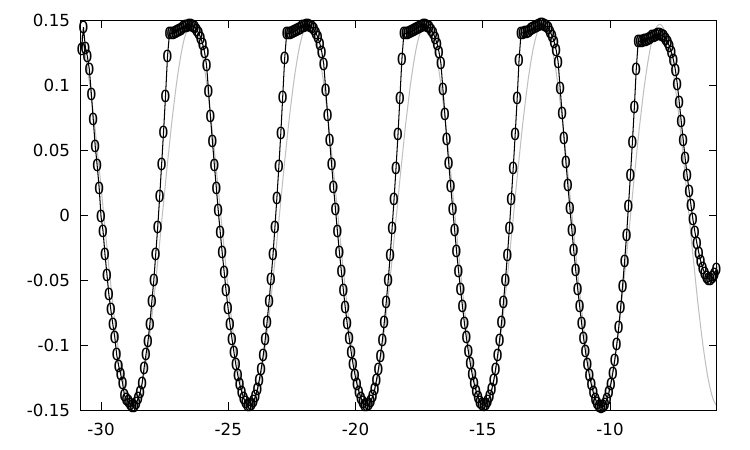} 
\caption{$\epsilon_2=0$: Numerical determination of the curvature
  scaling function $f_T$. Here and in all following plots of $f_T$,
  the horizontal axis shows $\ln(p_*-p)+B$ and the vertical axis shows
  $-1/2\ln T-A$. To make the plots larger, we omit axis labels in all
  line plots throughout. Circles indicate our data points, 30 per
  decade in $p-p_*$. We show only the range where the function is
  approximately periodic. The grey line represents a fit as a sine
  wave of amplitude $0.147$. For comparison, in all following plots of
  $f_T$ and $f_M$ the range of $\ln(p_*-p)+B$, will be the same as
  here, namely $[-30.8,-5.8]$.}
\label{fig:eps0_Tscaling}
\end{figure}

\begin{figure}
\includegraphics[scale=0.67, angle=0]{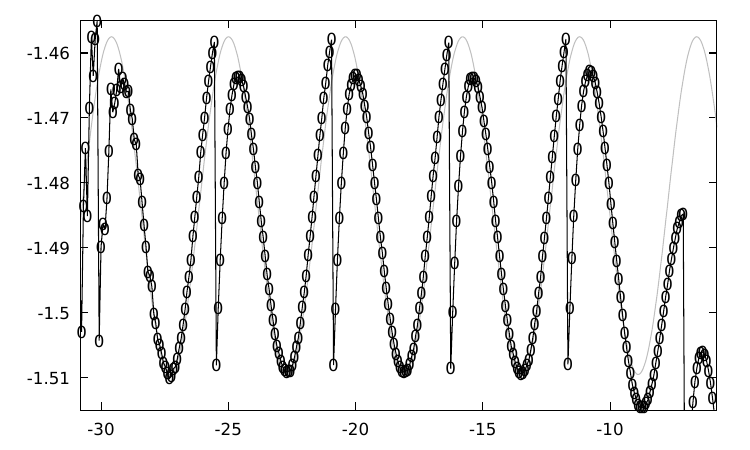} 
\caption{$\epsilon_2=0$: Numerical determination of the mass scaling
  function $f_M$. Here and in all following plots of $f_M$, the
  horizontal axis shows $\ln(p-p_*)+B$ and the vertical axis shows
  $\ln M-A$. The grey line represents a fit as $-1.4835$ plus a sine
  wave of amplitude $0.026$. $f_M$ is strictly periodic over a
  slightly smaller range than $f_M$.}
\label{fig:eps0_Mscaling}
\end{figure}

Our numerical measurement of $f_T$ with these fitting parameters is
shown in Fig.~\ref{fig:eps0_Tscaling}. We see accurate periodicity
from sub2.5 to sub13. The function $f_T$ is well approximated by a
fundamental sine wave of amplitude $0.147$, except for a widened top
and a clear discontinuity in the derivative at the left edge of this
flattened top. This kink is expected: it corresponds to the appearance
of a new local maximum of $T$ in each scale period.

Our numerical measurement of $f_M$ is shown in
Fig.~\ref{fig:eps0_Mscaling}. Recall that $M$ is the Hawking mass of
the first surface ${\cal S}_{u,x}$ with Hawking compactness
$C=0.99$. $f_M$ is periodic with an amplitude of about $0.026$ and
mean $-1.4835$. There is one large discontinuity in each period,
corresponding to the formation of the first MOTS one scale period
later. Again this is expected: when a new local maximum of $C$ takes
over, by definition it has the same $C$, but not the same $M$. We see
accurate periodicity with the same $\gamma$ and $\Delta$ as for the
curvature scaling law, over a slightly smaller range from sub4 to
sub12. The range may be smaller because the MOTS occurs near the past
light cone of the singularity of the underlying critical solution,
whereas the maximum of the Ricci curvature occurs at the centre, which
may be less affected by details of the initial data.

We are aware of the following previous plots of scaling law
fine-structures in the critical collapse of a spherical scalar field
in the literature. 

P\"urrer, Husa and Aichelburg \cite{PuerrerHusaAichelburg} extend
compactified Bondi coordinates to future null infinity, and so can
read off the asymptotic black hole mass. They show a periodic fine
structure of this mass which appears to be continuous, with an
amplitude of about $0.4$ in $\ln M$. Crespo, Oliveira and Winicour
\cite{CrespoOliveiraWinicour2019} use an affine radial coordinate also
compactified to null infinity and show a similar fine structure, but
with an amplitude that seems to be closer to $0.3$ in $\ln M$. These
results are in the same time slicing, so they should agree, and they
roughly do. From a plot in Rinne \cite{Rinne2020} we estimate the
amplitude in $\ln M$ as $\simeq 0.4$, similar again to
\cite{PuerrerHusaAichelburg,CrespoOliveiraWinicour2019}, although this
is not the asymptotic mass.  By contrast, our $f_M$ has an amplitude
of only $0.026$ in $\ln M$, and has a jump of similar amplitude.  Our
best guess is that the Hawking mass of our collapse diagnostic is
still very far from the final black hole mass. We have explained why
our $f_M$ is discontinous, but by contrast we have no theoretical
understanding of why the $M(p)$ measured by the other researchers is
continuous.

Switching now to the subcritical scaling of the maximum of the Ricci
curvature, Garfinkle and Duncan \cite{GarfinkleDuncan1998} plot the
fine-structure in $\ln\max|\text{Ric}|$, and from this plot we
estimate the amplitude of $f_T$ as $0.3$ in $\ln\max|\text{Ric}|$, or
$0.15$ in $(1/2)\ln\max|T|$. This is consistent with our value of
$0.147$. The bottom of their curve, corresponding to the top of ours,
is flattened, again consistent with our observation. Baumgarte
\cite{Baumgarte2018} has a fine-structure with amplitude of about $0.35$ in
$\ln\max\rho$, where $\rho=n^an^b T_{ab}$ is the energy density
measured by an observer normal to the time slicing. This depends on
the slicing, but the amplitude of $0.35$ is again similar to the $0.3$
observed by us and \cite{GarfinkleDuncan1998} for
$\ln\max|\text{Ric}|$.


\subsection{Self-similarity of the critical solution}


We demonstrate self-similarity of the critical solution by showing
that the scalar field $\psi$, the compactness $C$ and the scaled
curvature scalar
\begin{equation}
\label{calTdef}
{\cal T}:=(u_*-u)^2|\nabla_a\psi\nabla^a\psi|
\end{equation}
are periodic in the DSS-adapted coordinates
\begin{eqnarray}
\label{xidef}
\xi&:=&{R \over u_*-u}, \\
\label{taudef}
\tau&:=&-\ln(u_*-u).
\end{eqnarray}
The null slicing is already DSS-compatible, in the sense of
\cite{Corsetal2023}, but we call this specific coordinate system
DSS-adapted as the parameter $u_*$ needs to be adjusted to the
accumulation point of the specific DSS. Unlike the scalar field, the
metric actually has period $\Delta/2$ because the Choptuik solution
has the property that $\psi_*(\xi,\tau+\Delta/2)=-\psi_*(\xi,\tau)$
and $R_{ab}=8\pi\psi_{,a}\psi_{,b}$ is invariant under $\psi\to-\psi$.
The parameter $u_*$ is a family-dependent constant. A rough initial
guess of $u_*$ is given by the time that near-critical supercritical
evolutions stop. We then refine it by making $\psi$, $C$ and ${\cal
  T}$ as periodic as possible. For our family, we find $u_*\simeq
5.609$.

Note that adjusting $u_*$ slightly will affect $\tau(u)$ the more the
closer $u$ gets to $u_*$. In practice, we can therefore adjust the
$\tau$-{\em location} of the last echo of, say, $\psi(x,\tau)$ by adjusting
$u_*$, almost without moving the other echos. By contrast, this
arbitrariness is absent when we use $u_*$ to make the {\em amplitude}
of the last echo of ${\cal T}(x,\tau)$ equal to that of the preceding
echos, and is therefore what we use to determine $u_*$.

\begin{figure}
\includegraphics[scale=0.7, angle=0]{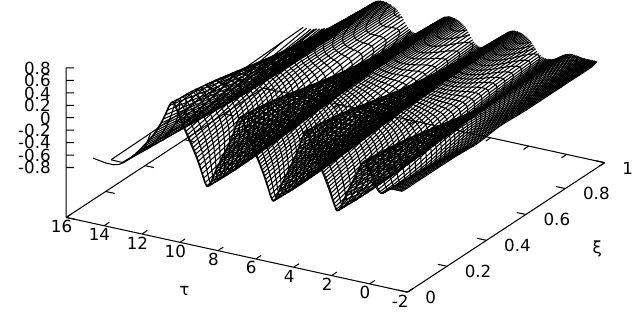} 
\caption{$\epsilon_2=0$: Surface plot of the scalar field
  $\psi(x,\tau)$ against the similarity coordinates $\xi$ and $\tau$,
  in our closest subcritical evolution (sub15). The plot has been
  cropped to $0\le \xi\le 1$, but the entire range of $\tau$ is shown,
  starting from the initial data at $u=0$ at the right edge of the
  plot, and ending at the left edge when $C\le 0.01$ indicates
  dispersion. The centre is at $\xi=0$ (front edge), and the past
  lightcone of the singularity at $\xi\simeq 1$ (back
  edge). Approximate DSS is seen for the range $2\lesssim \tau
  \lesssim 10$. Our numerical data form a mesh made up of lines of
  constant $u$ (constant $\tau$) and constant $x$ ({\em not} constant
  $\xi$). Both have been down-sampled for visual clarity.}
\label{fig:eps0_PsiDSS}
\end{figure}

\begin{figure}
\includegraphics[scale=0.7, angle=0]{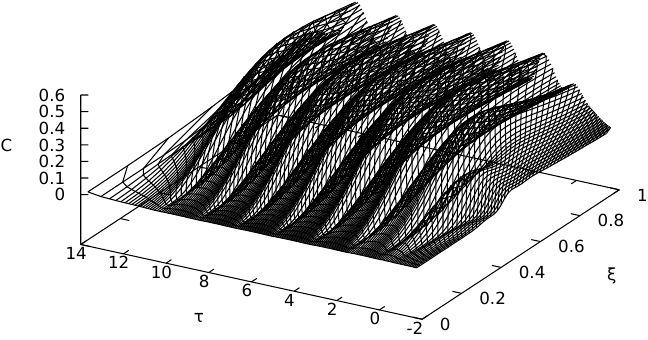} 
\caption{$\epsilon_2=0$: Surface plot of the compactness $C$, otherwise as
  described in Fig.~\ref{fig:eps0_PsiDSS}.}
\label{fig:eps0_CDSS}
\end{figure}

\begin{figure}
\includegraphics[scale=0.7, angle=0]{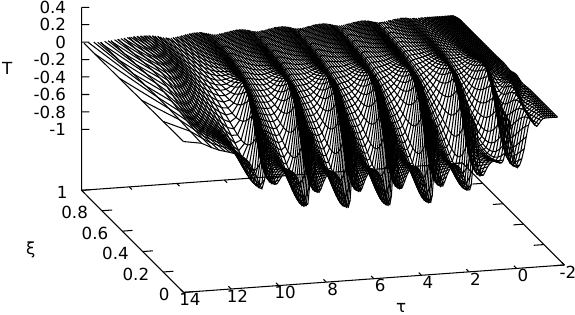} 
\caption{$\epsilon_2=0$: Surface plot of the scaled curvature
  diagnostic ${\cal T}$, otherwise as in Fig.~\ref{fig:eps0_PsiDSS},
  except that the plot has been rotated slightly to the right for a
  clearer view.}
\label{fig:eps0_TDSS}
\end{figure}

We use our closest super- and subcritical evolutions as a proxy for
the critical solution. They agree with each other until
one disperses and the other one forms a black hole. We see clear
echoing in $\psi$, $C$ and ${\cal T}$ for $2\lesssim \tau \lesssim
10$. The results are shown in Figs.~\ref{fig:eps0_PsiDSS},
\ref{fig:eps0_CDSS} and \ref{fig:eps0_TDSS} for our closest
subcritical evolution.

\subsection{Comparison with gsB gauge}
\label{section:gsBgauge}


We have also bisected $\epsilon_2=0$, $N_y=1$ in ``global shifted
Bondi'' (gsB) gauge. Recall from Paper~I that this
  is given by $R=s(u)x$, giving 
\begin{equation}
\left(B\right)={\dot s(u)x-\Xi R\over s(u)}, 
\end{equation}
and we have settled on the particular flavor given by
\begin{equation}
\dot s(u)={1\over x_0}\min_y\Xi R(u,x_0,y).
\end{equation}
 
We needed $x_\text{max}=20$ in gsB gauge (instead of $11$ in lsB
gauge), and correspondingly $N_x=800$ to maintain $\Delta
x=0.025$. Comparing the zoomed-in scaling laws with lsB and gsB, they agree
as expected down to about super/sub12, but then begin to disagree as
both become noisy.

In spite of this apparent success, we see a fundamental problem with
gsB gauge already here in spherical scalar field collapse. While $B$
is a monotonically decreasing function in weakly curved spacetimes,
going through $B=1$ at $x=0$ and $B=0$ at $x=x_0$, in near-critical
spacetimes it becomes non-monotonic and may cross zero at $x=x_0$ with
{\em positive} slope. This means that some surfaces of constant
$x>x_0$ are no longer future spacelike but timelike. This could
include the outer boundary, and we would then not evolve on the domain
of dependence. It is possible that $B$ becomes negative again for
sufficiently large $x$ , and indeed this happens for $x_\text{max}=20$
in this family of spherical initial data, but it is not something we
can rely on.

Given that the only free parameter of gsB gauge is $x_0$, and that
this needs to be set to $x_*$, which in turn is set by the family of
initial data, in order to resolve the critical solution this seems to
be a fundamental problem with gsB gauge. Indeed, gsB gauge fails for
this reason in nonspherical evolutions, and we present no results in
this gauge.


\section{Results for nonspherical initial data}
\label{section:nonspherical}


As already noted, the strong-field convergence tests with
$\epsilon_2=0.75$ in Appendix~\ref{appendix:convergencetestsold}
showed an initial burst of non-convergent error caused by our initial
data not being analytic at the origin, but we see second-order
convergence with $\Delta x$ and $l_\text{max}$ for all $u\gtrsim
0.2$. Here we present results for these non-analytic initial data, as
we believe the initial error does not affect our results in the
critical regime.

We have bisected in the families with $\epsilon_2=0$, $10^{-4}$,
$10^{-2}$, $0.1$, $0.5$ and $0.75$, using lsB4 gauge. Note our family
of null data parameterised by $(p,\epsilon_2)$ is equivalent to the
family of Cauchy data parameterised by $(p,\epsilon^2)$ of
\cite{Choptuiketal2003, Baumgarte2018} only in the limit of small
$\epsilon_2$ and small $p$, as we rely on a spherically symmetric
solution of the wave equation on flat spacetime to relate the scalar
$\psi$ field at $t=0$ to $u=0$. Table~\ref{table:epsruns} lists the
numerical parameters and outcomes of our successful bisections. To
estimate the error from discretization in $x$ and $y$, we have run
$\epsilon_2=0.75$ at four resolutions, with $\Delta x=0.05$ and
$0.025$, and $\bar N_y=9$ and $17$ for both. However, because of the
large number of long evolutions involved, we have not plotted the fine
structures of the scaling laws for the combination of the two higher
resolutions. Moreover, at higher fine-tuning during the bisection, we
had to reduce our criterion for collapse from $C\ge 0.99$ to $C\ge
0.8$. This would of course affect the mass scaling law (if we computed
it), but appears to be sufficient for the bisection, and hence finding
the approximate critical solution.

From (\ref{Phieps}), we expect that for sufficiently small
$\epsilon_2$, the non-sphericity evolves as a linear perturbation of
spherical symmetry, dominated by $l=2$, and that this linear
perturbation decays. For somewhat larger $\epsilon^2$,
\cite{Choptuiketal2003, Baumgarte2018} also found changes to $\Delta$
and $\gamma$. For even larger $\epsilon^2$, in particular
$\epsilon^2=0.75$, Baumgarte \cite{Baumgarte2018} also found evidence
of the (nonlinear) instability of the Choptuik solution reported by
Choptuik {\it et al} \cite{Choptuiketal2003} to lead to two centres of
collapse. We will see, by contrast, that for our $\epsilon_2=0.75$
evolutions the nonspherical perturbations appear to still be in the
linear regime, and indeed their non-sphericity is more similar to the
evolutions of Baumgarte \cite{Baumgarte2018} with his
$\epsilon^2=0.5$.

\begin{table*}
\setlength{\tabcolsep}{6pt} 
\renewcommand{\arraystretch}{1.3} 
\begin{tabular}{l|lllll|llll|l}
$\epsilon_2$ & $\Delta x$ & $\bar N_y$ & $l_\text{max}$ & $x_0$ &
  $x_\text{max}$ & $p_*$ & $u_*$ & $A$ & $B$ & comments \\
\hline
0 & 0.025 & 1/3 & 0/4 & 8.24 & 11. & 0.2402 & 5.609 &
1.113 & 0.592 & no filtering, Figs.~1-7, 12-13 \\
$10^{-4}$ & * & 9 & 8 & * & * & * & * & * & * & Figs.~12-13  \\
$10^{-2}$ & * & 3/5/9 & 2/4/8 & 8.275 & 15. & * & 5.610 & 1.115  & * 
&  Figs.~6-9, 12-15, 18-19, 22 \\
0.1 & * & 3/5/9 & 2/4/8 & 8.7 & * & 0.2406 & 5.620 & 1.131 &
* &  Figs.~6-7, 12-15, 18-19 \\
0.5 & * & 5/9 & 4/8 & 12.5/13.3 & * & 0.2452 &
5.657 & 1.230  & 0.737 &  * \\
\hline
0.75  & 0.05 & 9 & 8 & 24. & 50. & 0.2555 & 5.642 & 1.250 & 1.059
& Figs.~10-11\\
* & * & 17 & 16 &  30. & * & 0.2556 & * & * & * & * \\
* & 0.025 & 9 & 8 & 24.5 & * & 0.2535 & 5.651 & * & * & * \\
* & * & 17 & 16 & 33. & * & * & 5.652 & * & * &  
 $C\ge 0.8$, Figs.~6-7, 10-21, 23  \\
\end{tabular}
\caption{Table showing parameters of our families of initial
  data. $\epsilon_2$, in the first column, is the only physical input
  parameter. The second group group of columns shows numerical (input)
  parameters, and the third group physical (observed) parameters
  (fitted to the numerical data, and rounded to four significant
  digits for this table). These families, or the best subcritical
  evolutions in them, have been used in the figures listed in the last
  column. For visual clarity, a * means the entry is unchanged from
  the previous row. Half-frequency filtering was applied except in the
  first row, and the collapse criterion was $C\ge 0.99$ except in the
  last row, where it was $C\ge 0.8$.}
\label{table:epsruns}
\end{table*}


\subsection{The final collapse phase}
\label{section:almostsphericalfinalcollapse}


Our best supercritical evolution remains, by definition, close to our
best subcritical one until the end when one decides to disperse and
the other to collapse. However, in its final collapse phase the best
supercritical solution becomes much more non-spherical, and this is
challenging numerically.

In flat spacetime $\Xi R=-1/2$, $R\rho_+=1$ and $2R\rho_-=-1$. By
contrast, in critical collapse, the divergence $\rho_+$ of the
outgoing null geodesics remains approximately spherical and positive
in the self-similar phase, but in the final collapse becomes very
small at large $x$. Similarly, $\rho_-$ and $\Xi R$ remain negative
and approximately spherical in the self-similar phase, but become very
large, and positive for a range of angles intermediate between the
poles and equator, in the final collapse phase.

We also see a fundamental numerical challenge of highly nonspherical
evolutions in lsB gauge: The $x$-shift $B$ in
\begin{equation}
\bar R_{,u}=\Xi R+BR_{,x}
\end{equation}
tries to counteract the non-spherical part of $\Xi R$ to keep $R$
independent of $y$ but has little to act on as $R_{,x}\to 0$, while
$\Xi R$ also becomes larger. Therefore $B$ becomes large, and by the
Courant condition applied to the upwinded shift terms, the time step
becomes small. In spherical symmetry this is less of a problem because
$R_{,x}\to 0$ also indicates an approaching MOTS, whereas in a
non-spherical spacetime it does not.

Another consequence of the asphericity of the final collapse phase is
that we do {\em not} then have sufficient angular resolution to obtain
reliable values for $M$. In the continuum $\tilde C=C$ and $\tilde
M=M$, but in the final collapse phase their numerical values differ
strongly. There is little hope that we can overcome this problem with
more angular resolution, as physically we expect the ${\cal S}_{u,x}$
to become arbitrarily non-spherical inside the black hole. As a
heuristic indicator of collapse for the purpose of bisecting to the
black hole threshold, we therefore use $\tilde C$ rather than $C$, as
$\tilde M$ is at least non-decreasing by construction. Of course
either $\tilde M$ or $M$ should be considered reliable only to the
extent they agree with each other.


\subsection{Scaling and echoing}


We look for three types of evidence that the non-spherical
perturbations are in the linear regime: for sufficiently small
$\epsilon_2$, the spherical part of the critical solution should be
the same critical solution as for $\epsilon_2=0$, and its perturbations
should decay and oscillate as predicted by linear perturbation theory
about the spherical critical solution. The growth rate $\lambda_0>0$
of the unique growing mode should be unchanged, and hence the scaling
laws, including $\gamma$, $f_T$ and $f_M$ should be as for
$\epsilon_2=0$. Finally, the decay rates $\lambda$ of the leading
perturbation modes should be independent of $\epsilon_2$ for
$\epsilon_2$ sufficiently small, with the values predicted in
perturbation theory in \cite{Gundlach1997}. We examine evidence for
all this in the next two subsections, starting here with the scaling
laws.

\begin{figure}
\includegraphics[scale=0.67, angle=0]{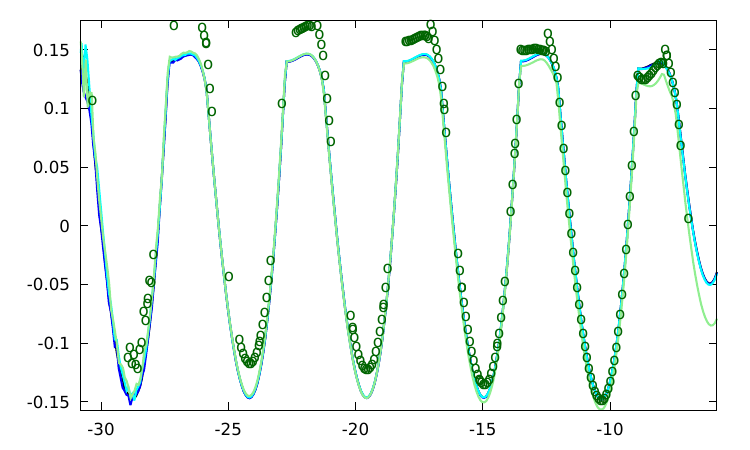}
\caption{Comparison of the curvature scaling functions $f_T$ for the
  $\epsilon_2=0$ (black line), $\epsilon_2=10^{-2}$ (blue line),
  $\epsilon_2=0.1$ (cyan line), $\epsilon_2=0.5$ (light green line)
  and $\epsilon_2=0.75$ (dark green circles) families. Each $f_T$ is
  computed at the highest numerical resolution shown in
  Table~\ref{table:epsruns}. Here and in the following plots, the
  horizontal range is exactly the same in
  Figs.~\ref{fig:eps0_Tscaling} and \ref{fig:eps0_Mscaling}, and the
  distribution of data points is the same as indicated by circles
  there, but for clarity we no longer show those individual data
  points, except for $\epsilon_2=0.75$, where our computation at the
  highest resolution has gaps due to lack of computing time.}
  \label{fig:compare_eps_Tscaling}
\end{figure}

\begin{figure}
\includegraphics[scale=0.67, angle=0]{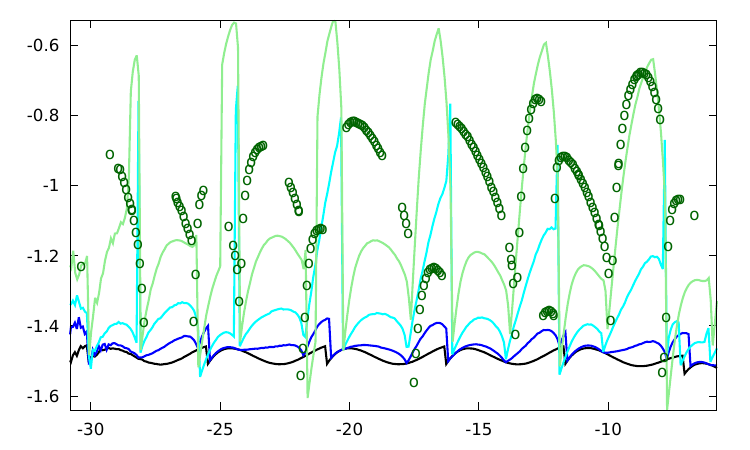}
\caption{Comparison of the mass scaling function $f_M$ for the same
  values of $\epsilon_2$ as in Fig.~\ref{fig:compare_eps_Tscaling}:
  $\epsilon_2=0$ (black), $\epsilon_2=10^{-2}$ (blue), $\epsilon_2=0.1$
  (cyan), $\epsilon_2=0.5$ (light green) and $\epsilon_2=0.75$ (dark
  green), at the same resolutions as there.}
  \label{fig:compare_eps_Mscaling}
\end{figure}

Fig.~\ref{fig:compare_eps_Tscaling} and
Fig.~\ref{fig:compare_eps_Mscaling} compare the fine structures of
$T^{-1/2}$ and $M$ scaling, respectively, for different $\epsilon_2$,
using our highest resolution data at each $\epsilon_2$. The same
$\gamma=0.374$ power law has been taken out for all values of
$\epsilon_2$. We are treating each value of $\epsilon_2$ as a
different 1-parameter family of initial data for the purpose of
fitting $A$ and $B$.

\begin{figure}
\includegraphics[scale=0.67, angle=0]{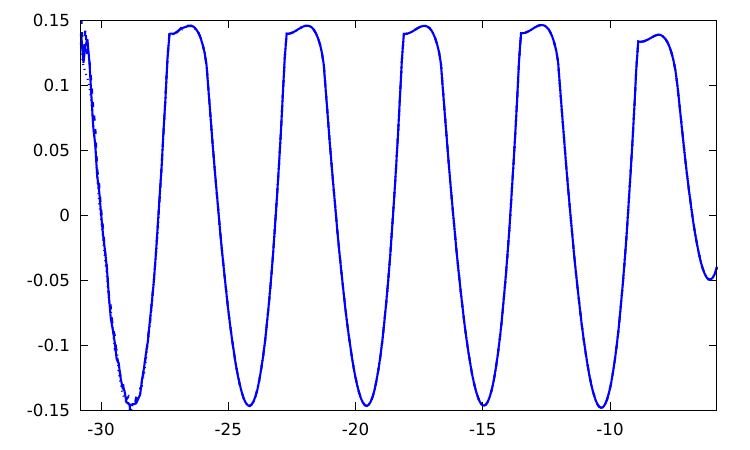} 
\caption{The curvature scaling function $f_T$ for $\epsilon_2=10^{-2}$
  (blue), at three different angular resolutions: $\bar N_y=3$ (solid)
  $\bar N_y=5$ (dashed) and $\bar N_y=9$ (dotted), and
  $l_\text{max}=2,4,8$, respectively, all with $\Delta x=0.025$,
  $x_0=8.275$, $x_\text{max}=15$. The curves are indistinguishable,
  indicating that the errors from discretizing in $y$ are
  small.}
\label{fig:eps1dm2_Tscaling_convergence}
\end{figure}

To estimate the accuracy of our scaling results for non-spherical
initial data, Figs.~\ref{fig:eps1dm2_Tscaling_convergence} and
Fig.~\ref{fig:eps0.75_Tscaling_convergence} show $f_T$ at different
resolutions for $\epsilon_2=10^{-2}$ and $\epsilon_2=0.75$,
respectively.  At $\epsilon_2=10^{-2}$ we have very good accuracy for
$f_T$. At $\epsilon_2=0.75$ we have some discretization error in the
amplitude of $f_T$, but essentially none in its period.

\begin{figure}
\includegraphics[scale=0.67, angle=0]{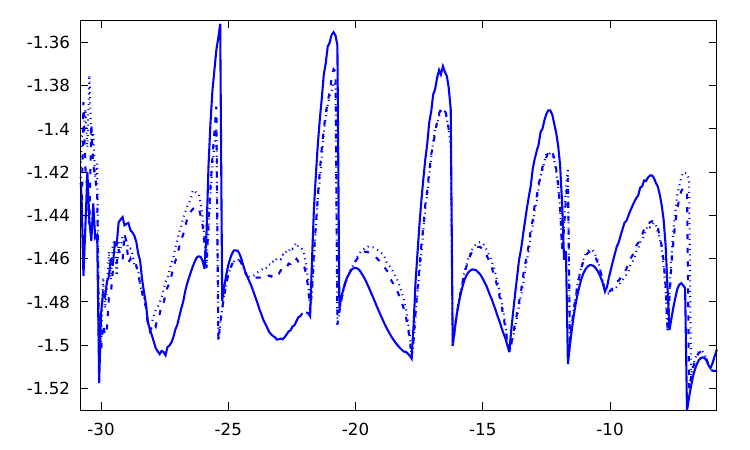} 
\caption{The mass scaling function $f_M$ for $\epsilon_2=10^{-2}$, at
  the same four resolutions as in
  Fig.~\ref{fig:eps1dm2_Tscaling_convergence}. The differences in
  $f_M$ between resolutions are very large, compared to $f_T$ in
  Fig.~\ref{fig:eps1dm2_Tscaling_convergence}.}
\label{fig:eps1dm2_Mscaling_convergence}
\end{figure}

\begin{figure}
\includegraphics[scale=0.67, angle=0]{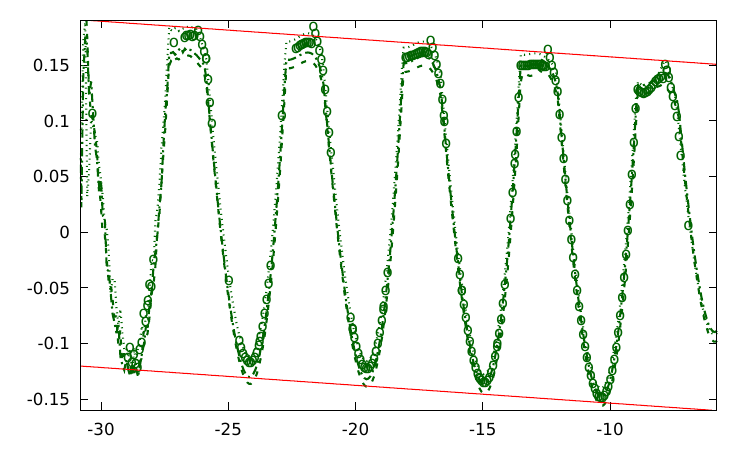} 
\caption{The curvature scaling function $f_T$ for $\epsilon_2=0.75$,
  at four different resolutions: $\bar N_y=17$, $\Delta x=0.005$ (our
  best resolution, circles), $\bar N_y=9$, $\Delta x=0.0025$,
  $x_0=24.5$ (dashed lines), and $\bar N_y=17$, $\Delta x=0.005$,
  $x_0=30$ (dotted), and $\tilde N_y=9$, $\Delta x=0.005$, $x_0=24$
  (dot-dashed). The red straight lines represent a fit by eye to the
  local minima and maxima of $f_T$. They have slope
  $\delta\gamma=-0.0016$.}
\label{fig:eps0.75_Tscaling_convergence}
\end{figure}

\begin{figure}
\includegraphics[scale=0.67, angle=0]{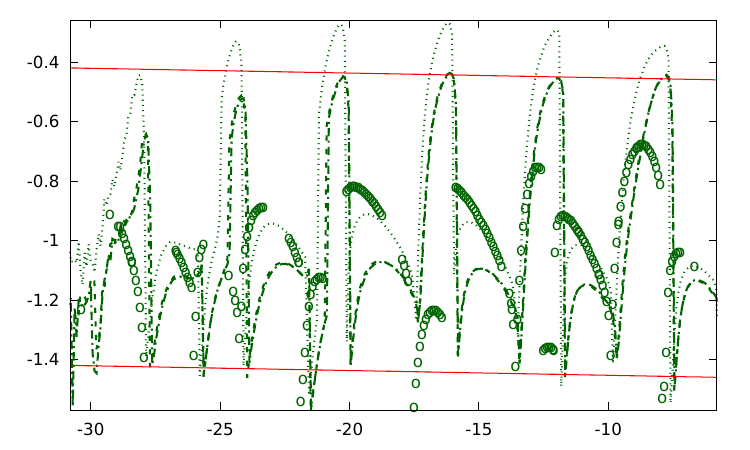} 
\caption{The mass scaling function $f_M$ for $\epsilon_2=0.75$, at the
  same four resolutions as in
  Fig.~\ref{fig:eps0.75_Tscaling_convergence}.  The red straight lines
  have slope $\delta\gamma=-0.0016$, taken from the fit in
  Fig.~\ref{fig:eps0.75_Tscaling_convergence}. $f_M$ is both not
  periodic enough and too resolution-dependent for a meaningful fit to
  its average slope, but $\delta\gamma=-0.0016$ is consistent with the
  data.}
\label{fig:eps0.75_Mscaling_convergence}
\end{figure}

On the other hand, Figs.~\ref{fig:eps1dm2_Mscaling_convergence} and
Fig.~\ref{fig:eps0.75_Mscaling_convergence} show significant
resolution dependence in the numerically measured $f_M$ already at
$\epsilon_2=10^{-2}$, and similar, but not much larger, at
$\epsilon_2=0.75$. We believe the variation of $f_M$ with $\epsilon_2$
shown in Fig.~\ref{fig:compare_eps_Mscaling} is real but, we do not
have enough numerical resolution for quantitatively reliable results.

Our plots of $f_T$ in Fig.~\ref{fig:compare_eps_Tscaling} at different
values of $\epsilon_2$ can be compared directly with Fig.~3 of
\cite{Choptuiketal2003} and Fig.~7 of \cite{Baumgarte2018}, with the
differences that we plot $-1/2\ln T+A$ rather than $\ln T$ on the
vertical axis, and $\ln(p_*-p)+B$ rather $\ln(p_*-p)$ on the
horizontal axis. In other words, \cite{Choptuiketal2003} and
\cite{Baumgarte2018} do not fit our family-dependent parameters $A$
and $B$, but in effect set them to zero. 

Setting that aside, our results show only a tiny effect of
$\epsilon_2$ on $\gamma$, compared to
\cite{Choptuiketal2003,Baumgarte2018,Maroudaetal2024}. In our most
non-spherical evolutions, $\epsilon_2=0.75$, we only see a change of
$\delta\gamma\simeq -0.0016$. By comparison, \cite{Choptuiketal2003,
  Baumgarte2018} find $\delta\gamma=-0.007$ and $-0.005$,
respectively, at their $\epsilon^2=0.5$, which appears to be the
nearest comparator to our $\epsilon_2=0.75$ (see also
Table~\ref{table:allcomparison}).
 
\begin{figure}
\includegraphics[scale=0.67, angle=0]{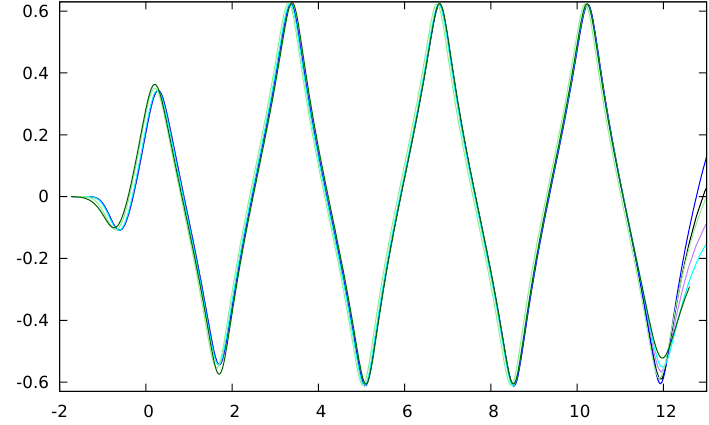} 
\caption{The scalar field at the centre, $\psi_0(0,\tau)$, at our best
  numerical resolution and best subcritical value of $p$, for each of
  $\epsilon_2=0$ (black), $10^{-4}$ (purple), $10^{-2}$ (blue), $0.1$
  (cyan), $0.5$ (light-green) and $0.75$ (dark-green). The horizontal
  axis shows $\tau$ and the vertical axis $\psi$ at the centre.}
\label{fig:eps_psicentre}
\end{figure}

\begin{figure}
\includegraphics[scale=0.67, angle=0]{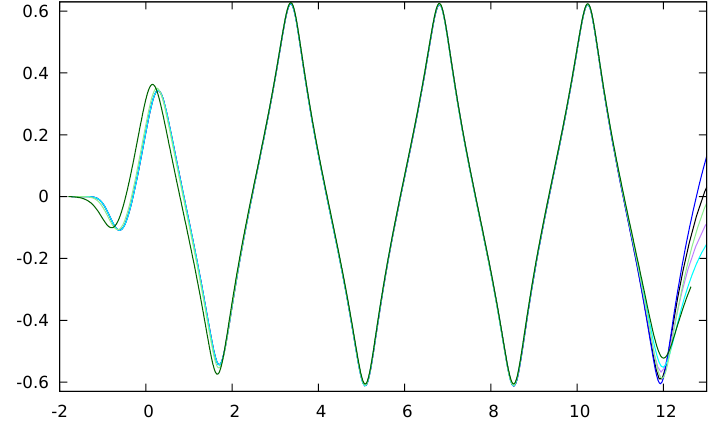} 
\caption{The same plot as in in Fig.~\ref{fig:eps_psicentre}, but a
  linear transformation applied to $\tau$ applied to $\epsilon_2=0.1$,
  $0.5$ and $0.75$ (but not the smaller values), in order to align the
  first and third full maxima. For this, $\tau$ is stretched by
  factors of $0.9995$, $1.0008$ and $1.0068$, respectively, as well as
  shifted.}
\label{fig:eps_psicentre_shifted_stretched}
\end{figure}

We also see only a tiny deviation of the period of $f_T$ from its
theoretical value $\Delta/(2\gamma)$ (with $\Delta=3.44$ and
$\gamma=0.374$) for all values $\epsilon_2=0...0.75$.  This is
demonstrated most clearly in Figs.~\ref{fig:eps_psicentre} and
\ref{fig:eps_psicentre_shifted_stretched}, which show $\psi(0,\tau)$
(technically, $\psi_0$ extrapolated to $x=0$, which is not on the
grid). Our best fits are $\delta\ln\Delta\simeq 5\cdot 10^{-4}$,
$-8\cdot 10^{-4}$ and $-64\cdot 10^{-4}$ at $\epsilon_2=0.1$, $0.5$
and $0.75$. These are not even monotonic in $\epsilon_2$ and therefore
more likely due to numerical errors than a physical change.

For more indirect evidence of $\Delta$ from the periodicity of fine
structure of the scaling laws, see also Fig.~\ref{fig:eps0_Tscaling}
for a fit of $f_T$ at $\epsilon_2=0$ to a sine wave of period $3.44$, and
Fig.~\ref{fig:compare_eps_Tscaling} for a comparison of $f_T$ 
with values from $\epsilon_2=0$ up to $\epsilon_2=0.75$.

We conclude that $\Delta$ does not change by more than $10^{-3}$ up to
our $\epsilon_2=0.75$. By comparison, \cite{Choptuiketal2003,
  Baumgarte2018} find $\delta\Delta=-0.05...-0.12$ at their
$\epsilon^2=0.5$ (see also Table~\ref{table:allcomparison}).

Unfortunately, \cite{Choptuiketal2003} and \cite{Baumgarte2018} do not
present plots of $f_M$. Our own results in
Fig.~\ref{fig:compare_eps_Tscaling} show that $f_M$ differs
significantly from $\epsilon_2=0$ already at $\epsilon_2=10^{-2}$, and
has become very much larger at $\epsilon_2=0.75$. Recall however that
our $M$ is only a very rough indication of the final black hole mass,
particularly so in our best resolution for $\epsilon_2=0.75$ where we
have measured the area of a two-surface with Hawking compactness of
only $0.8$. We therefore do not believe our results for $f_M$ are
accurate, in contrast to our results for $f_T$.

\subsection{Self-similarity and evolution of the nonsphericity}


We now test the hypothesis that there is a regime of small
$\epsilon_2$ where near-critical time evolutions can be approximated
by the spherical critical solution plus small perturbations, and
attempt to find the limit of its validity.

We begin with the spherical part of the scalar field and metric.
Under the assumption that ${\cal T}$ is still dominated by the
derivatives of $\psi_0$, we adjust $u_*$ to make ${\cal T}$ as
periodic as possible, and in particular with as many maxima and minima
taking the same values, just as in the spherical case. We then find
that $\psi_0(x,\tau)$, and $C(x,\tau)$, with the same fitted value of
$u_*$, are essentially identical from $\epsilon_2=0$ through to
$\epsilon_2=0.75$. See again the limit on a possible change of
$\Delta$ of the critical solution documented above.

We next address the hypothesis that the deviations from spherical
symmetry are small and essentially linear.

To find the linear perturbations of a given spherical DSS critical
solution $\phi_*$, one can separate the linearised equations by
angular dependence $l$, and for each $l$ and dimensionless field
$\phi$ make a mode ansatz
\begin{equation}
\delta\phi_l(\xi,\tau)=e^{\lambda \tau}\,\delta\tilde\phi_{l\lambda}(\xi,\tau),
\end{equation}
with $\delta\tilde\phi_{l\lambda}$ defined to be periodic in $\tau$
with period $\Delta$, and regular at the centre and at the past
lightcone.  Here $\phi$ stands for any scale-invariant quantity, such
as $\psi$ or ${\cal T}$. From this ansatz, the complex mode function
$\delta\tilde\phi_{l\lambda}$ and corresponding complex Lyapunov
exponent
\begin{equation}
\lambda:=\kappa+i\omega
\end{equation}
are then determined as the eigenfunctions and eigenvalues of a
(singular) linear boundary value problem \cite{JMMGundlach1999}. As
the background solution and its perturbations are real (the complex
mode ansatz is only for convenience), $\lambda$ and the corresponding
$\delta\tilde\phi_{l\lambda}$ are either real or form complex
conjugate pairs.

From these complex modes, one can construct the corresponding real
perturbations as
\begin{eqnarray}
\delta\phi_l(\xi,\tau)&=&{\rm Re} \left[Ae^{i\alpha} e^{(\kappa+i\omega)\tau}
  \,\delta\tilde\phi_{l\lambda}(\xi,\tau)\right] \nonumber \\ 
&=&Ae^{\kappa \tau}[\cos(\omega
  \tau+\alpha)\, {\rm Re}\, \delta\tilde\phi_{l\lambda}(\xi,\tau) \nonumber \\ 
&&-\sin(\omega \tau+\alpha)\,{\rm Im}\, \delta\tilde\phi_{l\lambda}(\xi,\tau)],  
\label{singlemodereal}
\end{eqnarray}
for arbitrary positive real amplitude $A$ and phase $0\le\alpha<2\pi$.  (Note
this is slightly incorrect in Eq.~(25) of Ref.~\cite{Baumgarte2018}).
Unless $\omega$ is a rational multiple of $2\pi/\Delta$, the product
$e^{-\kappa\tau}\delta\phi_l(\xi,\tau)$, while not growing or decaying, is
nevertheless not periodic in $\tau$ but only quasiperiodic.

The perturbation modes of the spherical scalar field critical solution
were found numerically in \cite{JMMGundlach1999}, using different
similarity coordinates from the ones defined here. We can therefore
compare the $\lambda$ with \cite{JMMGundlach1999}, but not directly
the $\delta\tilde\phi_{l\lambda}(\tau,x)$. As expected for a critical
solution, \cite{JMMGundlach1999} finds a single growing $l=0$ mode,
with $\lambda$ and the corresponding mode function real. All other
spherical modes, and all non-spherical modes, are complex and decay,
that is, $\kappa<0$. The least damped (most slowly decaying)
non-spherical mode was found to have $l=2$ angular dependence, with
$\kappa\simeq -0.07/\Delta$ and $\omega/(2\pi)\simeq
0.3/\Delta$. 

We expect that initial data which are almost spherical and fine-tuned
to the threshold of collapse, but otherwise generic, evolve into
something that can be approximated by the spherical critical solution
plus a linear perturbation, and that the perturbation can be
represented as sum of modes each with its own complex amplitude
$Ae^{i\alpha}$ determined by the initial data.

During the intermediate range of $\tau$ where the solution is
approximated by the critical solution plus small perturbations, we
expect the least damped $l=2$ perturbation to dominate the $l=2$
component of the solution. The single-mode formula
(\ref{singlemodereal}) should then approximately describe the $l=2$
component of near-critical evolutions in this regime. Moreover, its
amplitude should be proportional to $\epsilon_2$, as the $l=2$
component of the initial data is proportional to $\epsilon_2$ to
leading order.

In our initial data, the non-spherical part of $\psi$ is, to leading
order in $\epsilon_2$, purely $l=2$, and so, going to quadratic order,
we expect the initial data for the $l=4$ spherical harmonic
component $\psi_4$ to be proportional to $\epsilon_2^2$. During the
time evolution $\psi_4$ is sourced by terms that are linear in $f_4$
or quadratic in $f_2$ and $\psi_2$. $f_4$ is initially zero and then
sourced by terms linear in $\psi_4$ and quadratic in $f_2$ and
$\psi_2$. Perturbatively in $\epsilon_2$, we therefore expect all
$l=2$ components to be proportional to $\epsilon_2e^{\kappa\tau}$, all
$l=4$ components to be proportional to $\epsilon_2^2e^{2\kappa\tau}$,
and so on for higher $l$.

\begin{figure}
\includegraphics[scale=0.67, angle=0]{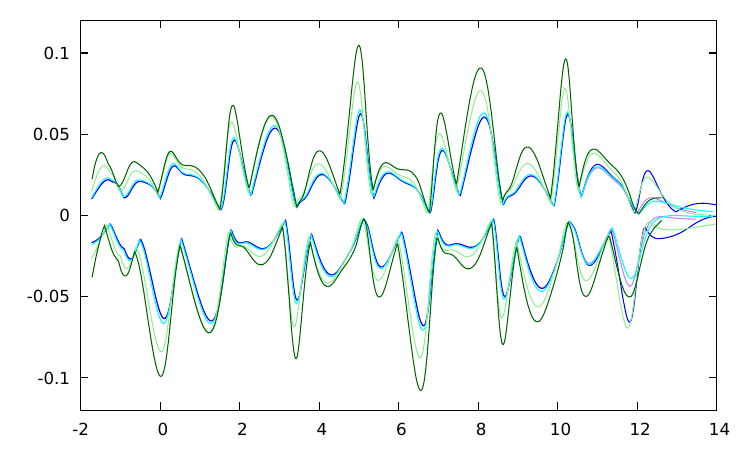}
\caption{ $\min_x$ and $\max_x$ of
  $\epsilon_2^{-1}e^{-\kappa\tau}\psi_2(x,\tau)$, plotted against $\tau$, for
  $\epsilon_2=10^{-4}$ (purple), $\epsilon_2=10^{-2}$ (blue),
  $\epsilon_2=0.1$ (cyan), $\epsilon_2=0.5$ (light green) and
  $\epsilon_2=0.75$ (dark green).}
\label{fig:psi2max}
\end{figure}

\begin{figure}
\includegraphics[scale=0.67, angle=0]{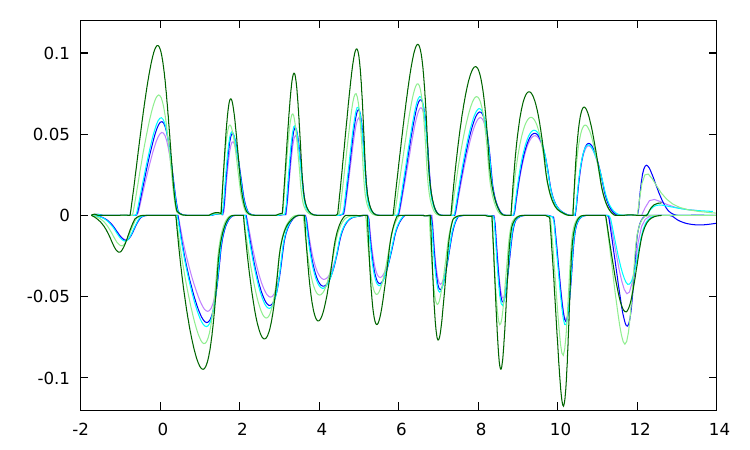}
\caption{ $\min_x$ and $\max_x$ of
  $\epsilon_2^{-1}e^{-\kappa\tau}f_2(x,\tau)$, plotted against
  $\tau$, otherwise as in Fig.~\ref{fig:psi2max}.}
\label{fig:f2max}
\end{figure}

To demonstrate the expected scaling with both $\tau$ and and
$\epsilon_2$, in Fig.~\ref{fig:psi2max} we plot the maximum and
minimum of $\epsilon_2^{-1}e^{-\kappa\tau}\psi_2(x,\tau)$ over $x$ in
the range $0\le x\le 1$ against $\tau$, for different $\epsilon_2$,
with our best subcritical $p$ at our best numerical resolution. The
same is done in Fig.~\ref{fig:f2max} for $f_2$. We find that
$\epsilon_2^{-1}e^{-\kappa\tau}\psi_2$ is essentially the same at
$\epsilon_2=10^{-4}$, $10^{-2}$ and $0.1$, and so is
$\epsilon_2^{-1}e^{-\kappa\tau}f_2$. At $\epsilon_2=0.5$ and $0.75$,
they are a bit larger (by a factor of $\sim 1.5$ at
$\epsilon_2=0.75$), but still agree qualitatively feature by
feature. In other words, $\psi_2$ and $f_2$ are almost linear up to
$\epsilon_2=0.1$, and still approximately linear up to $0.75$.

In these plots, we have fitted $\kappa=-0.03$ by eye. Compare this
with the value of $\kappa\simeq -0.07/\Delta\simeq -0.02$ determined
in \cite{JMMGundlach1999} from a numerical construction of the
eigenvalues of the operator evolving a linear perturbation from $\tau$
to $\tau+\Delta$. Neither is very accurate, but we believe that our
fitted value is compatible with the theoretically expected one, but
not with zero. In other words, we find that $\psi_2$ and $f_2$ decay
as expected from the linear perturbation mode analysis, up to
$\epsilon_2=0.75$.

\begin{figure}
\includegraphics[scale=0.8, angle=0]{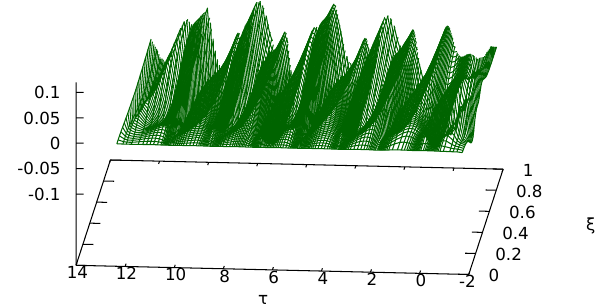}
\caption{Surface plot of the scaled non-spherical scalar field
  component $\epsilon_2^{-1}e^{-\kappa\tau}\psi_2(x,\tau)$, in the
  sub15 evolution of the $\epsilon_2=0.75$ family. Otherwise as
  described in Fig.~\ref{fig:eps0_PsiDSS}. Note that the dark green
  curves in Fig. \ref{fig:psi2max} are the projection of this surface
  plot in the $\xi$ direction.}
\label{fig:eps075_psi2}
\end{figure}

\begin{figure}
\includegraphics[scale=0.8, angle=0]{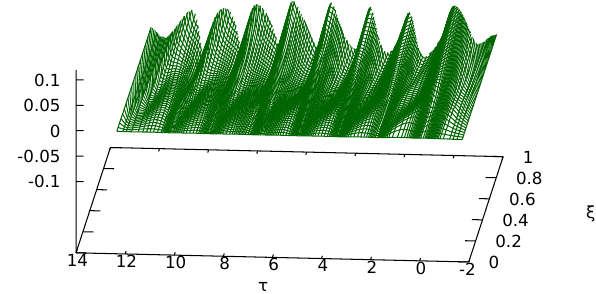}
\caption{Surface plot of the scaled metric component
  $\epsilon_2^{-1}e^{-\kappa\tau}f_2(x,\tau)$ , otherwise as in
  Fig.~\ref{fig:eps075_psi2}.}
\label{fig:eps075_f2}
\end{figure}

Having established these scalings, we give the full surface plots for
$\psi_2(x,\tau)$ and also $f_2(x,\tau)$, for $\epsilon_2=0.75$ in
Figs.~\ref{fig:eps075_psi2} and \ref{fig:eps075_f2}.

\begin{figure}
\includegraphics[scale=0.67, angle=0]{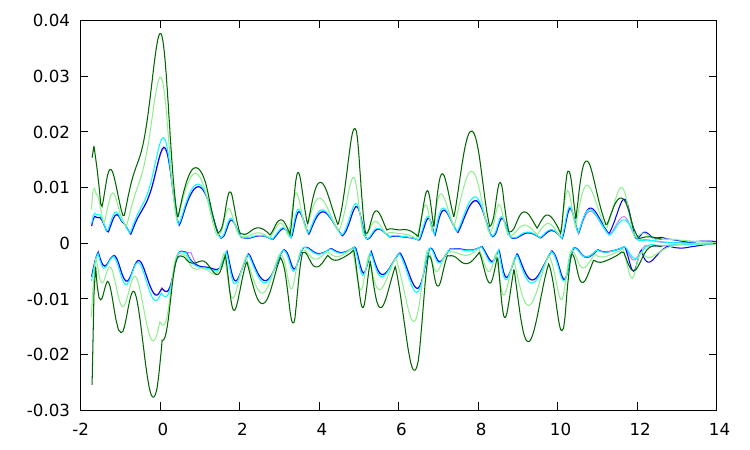}
\caption{ $\min_x$ and $\max_x$ of
  $\epsilon_2^{-2}e^{-2\kappa\tau}\psi_4(x,\tau)$, plotted against
  $\tau$, otherwise as in Fig.~\ref{fig:psi2max}.}
\label{fig:psi4max}
\end{figure}

\begin{figure}
\includegraphics[scale=0.67, angle=0]{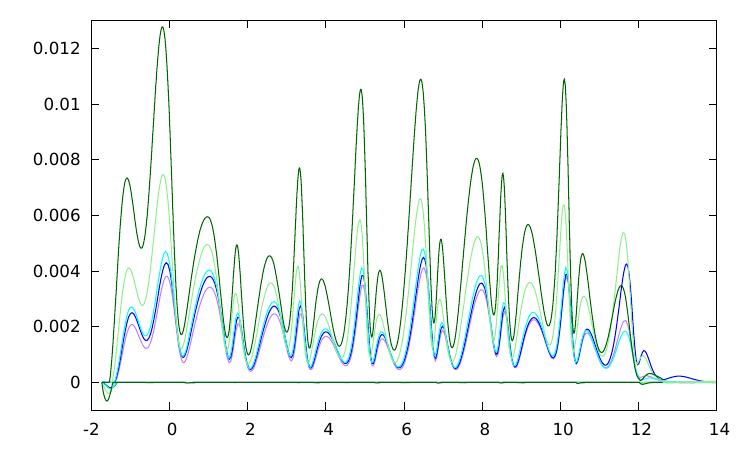}
\caption{ $\min_x$ and $\max_x$ of
  $\epsilon_2^{-2}e^{-2\kappa\tau}f_4(x,\tau)$, plotted against
  $\tau$, otherwise as in Fig.~\ref{fig:psi2max}. Where the minimum is
exactly zero, $f_4\ge 0$, with equality only at the origin.}
\label{fig:f4max}
\end{figure}

\begin{figure}
\includegraphics[scale=0.8, angle=0]{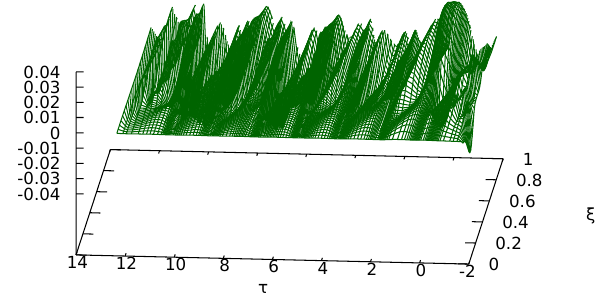}
\caption{Surface plot of the scaled non-spherical scalar field
  component $\epsilon_2^{-2}e^{-2\kappa\tau}\psi_4(x,\tau)$ against the
  similarity coordinates $\xi$ and $\tau$, in the sub15 evolution of
  the $\epsilon_2=0.75$ family.}
\label{fig:eps075_psi4}
\end{figure}

\begin{figure}
\includegraphics[scale=0.8, angle=0]{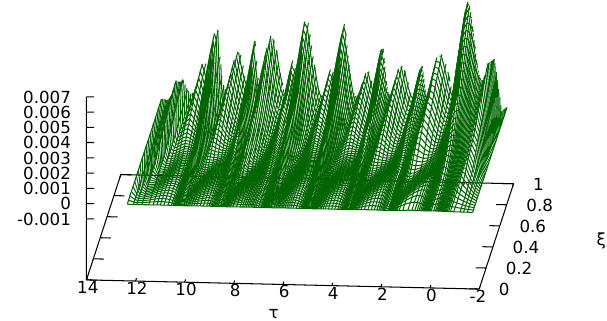}
\caption{Surface plot of the scaled metric component
  $\epsilon_2^{-2}e^{-2\kappa\tau}f_4(x,\tau)$ , otherwise as in
  Fig.~\ref{fig:eps075_psi4}.}
\label{fig:eps075_f4}
\end{figure}

The $l=4$ modes of $\psi$ and $f$ in the best subcritical evolutions
also seem to be quasiperiodic in $\tau$ and roughly independent of
$\epsilon_2$ when rescaled with
$\epsilon_2^{-2}e^{-2\kappa\tau}$. This is demonstrated in
Figs.~\ref{fig:psi4max}-\ref{fig:eps075_f4}.

The mode frequency $\omega$ could in principle be obtained by a
(discrete) Fourier transform of our data in $\tau$. We have not attempted
this as the region where the background spherical solution is clearly
DSS and the different perturbation amplitudes agree (after rescaling)
is only $1\lesssim \tau\lesssim 9$, giving us only about two periods of
the background solution. The discrete set of angular frequencies
present in the Fourier transform with respect to $\tau$ of
(\ref{singlemodereal}) is
\begin{equation}
\omega_N:=\omega+N{2\pi\over\Delta},\qquad N\in{\Bbb Z},
\end{equation}
where the cited $\omega$ is the value for $N=0$, that is, the smallest
positive frequency in the spectrum. Because of the symmetry
$\psi(x,\tau+\Delta/2)=-\psi(x,\tau)$, $N$ takes odd values for the
scalar field and even values for the metric. The spectrum obviously
depends on the value of the parameter $\alpha$ in
(\ref{singlemodereal}), but for a random $\alpha$,
\cite{JMMGundlach1999} found that the peak of the spectrum was at
$N=5$. The corresponding period in $\tau$ is
\begin{equation}
P_N:={2\pi\over|\omega_N|}={\Delta \over \left|N+{\omega\Delta\over 2\pi}\right|}
\simeq {3.44\over |N+0.3|}.
\end{equation}
If we count the peaks in Fig.~\ref{fig:psi2max}, we find about 12 in
the range $0\le\tau\le 12$, giving us a ``pseudoperiod'' of about 1,
similar to $P_3\simeq 1.0$. If we count the peaks in
Fig.~\ref{fig:f2max}, with find about 8 in the same range, giving us a
pseudoperiod of about $2/3$, similar to $P_5\simeq 0.65$. 

The numerical measurement of both $\kappa$ and $\omega$ would be
improved in proportion to increasing the range of $\tau$ over which
our fine-tuned numerical solutions approximate the critical
solution. Going any further in this would require quadruple precision,
not only so that $p$ can be represented to more significant figures,
but more importantly so that round-off error during the numerical
solution is suppressed better. 

We note in passing that $\psi_2/x^2$, $f_2/x^2$ and $f_4/x^4$, which should be
regular at $x=0$ in the continuum, are also numerically regular at
$x=0$, but $\psi_4/x^4$ is not. (It is generally hard to enforce
$f_l\sim r^l$ in any freely evolved variable, without explicitly
taking out the factor $r^l$.)


\subsection{Comparison with the evolutions of Choptuik et al,
  Baumgarte, and Marouda et al}


In order to compare the non-sphericity of our evolutions with those of
\cite{Baumgarte2018} (and by implication also of
\cite{Choptuiketal2003} for the same initial data), we note that
Baumgarte has measured the difference in $\psi$ on outgoing null
geodesics at $y=\pm 1$ and $y=0$ (poles and equator). Fig.~11 of
\cite{Baumgarte2018} shows this $\delta\psi$ for $\epsilon_2=10^{-2}$
during the approximately self-similar phase, plotted against the
similarity variables $\tau=-\ln(u_*-u)$ and
$\xi_\lambda:=\lambda/(u_*-u)$, where $\lambda$ is the affine
parameter, normalised to $\lambda=0$ and $d\lambda/dR=1$ at the
origin, and $u_*$ is fitted as described above. This measure is
gauge-invariant.

We note that for infinitesimal deviations from spherical symmmetry,
and assuming the deviation only has an $l=2$ component,
\begin{eqnarray}
\delta\psi(u,x)&:=&\psi(u,x,\pm
1)-\psi(u,x,0) \\
&=&\left[P_2(\pm1)-P_2(0)\right]\psi_2(u,x) \\
&=&{3\over 2}\psi_2(u,x), 
\end{eqnarray}
where $\psi_2$ denotes the $l=2$ component of $\psi$. $\lambda$ along
outgoing null geodesics is given by
\begin{equation}
\lambda(u,x,y)=\int_0^x G(u,x',y)\,dx'
\end{equation}

\begin{figure}
\includegraphics[scale=0.5, angle=0]{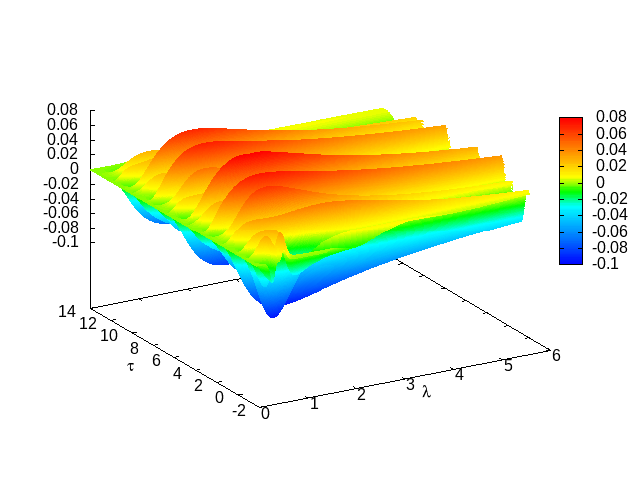}
\caption{Surface plot of $(3/2)/\epsilon_2\psi_2\simeq 
    \delta\psi/\epsilon_2$ for $\epsilon_2=10^{-2}$ against
  similarity coordinates $\tau$ and $\xi_\lambda$. This should be
  compared to Fig.~11 of \cite{Baumgarte2018}. The colour map has been
  chosen purely for visual agreement with that figure. The range of
  $\xi_\lambda$ is the same as there, but our range of $\tau$ is
  larger, $[0,14]$ rather than $[0,8]$, as we can fine-tune more
  closely to the black hole threshold.}
\label{fig:eps1dm2_psi2_lambda}
\end{figure}

The equivalent of Fig.~11 of \cite{Baumgarte2018} created from our
data for $\epsilon_2=10^{-2}$ is
Fig.~\ref{fig:eps1dm2_psi2_lambda}. Note that in this plot we have
{\em not} applied the factor of $e^{-\kappa\tau}$. (This makes little
difference, as $\kappa$ is so small.) At small $\epsilon_2$ our
solutions are a little more nonspherical than those of
\cite{Baumgarte2018}, but recall that we set initial data on different
slices.

\begin{figure}
\includegraphics[scale=0.5, angle=0]{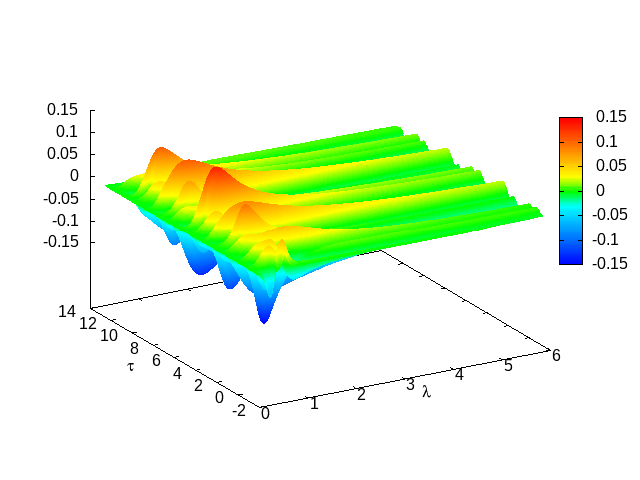}
\caption{As in Fig.~\ref{fig:eps1dm2_psi2_lambda}, but for $\epsilon_2=0.75$}
\label{fig:eps075_psi2_lambda}
\end{figure}

\begin{figure}
\includegraphics[scale=0.8, angle=0]{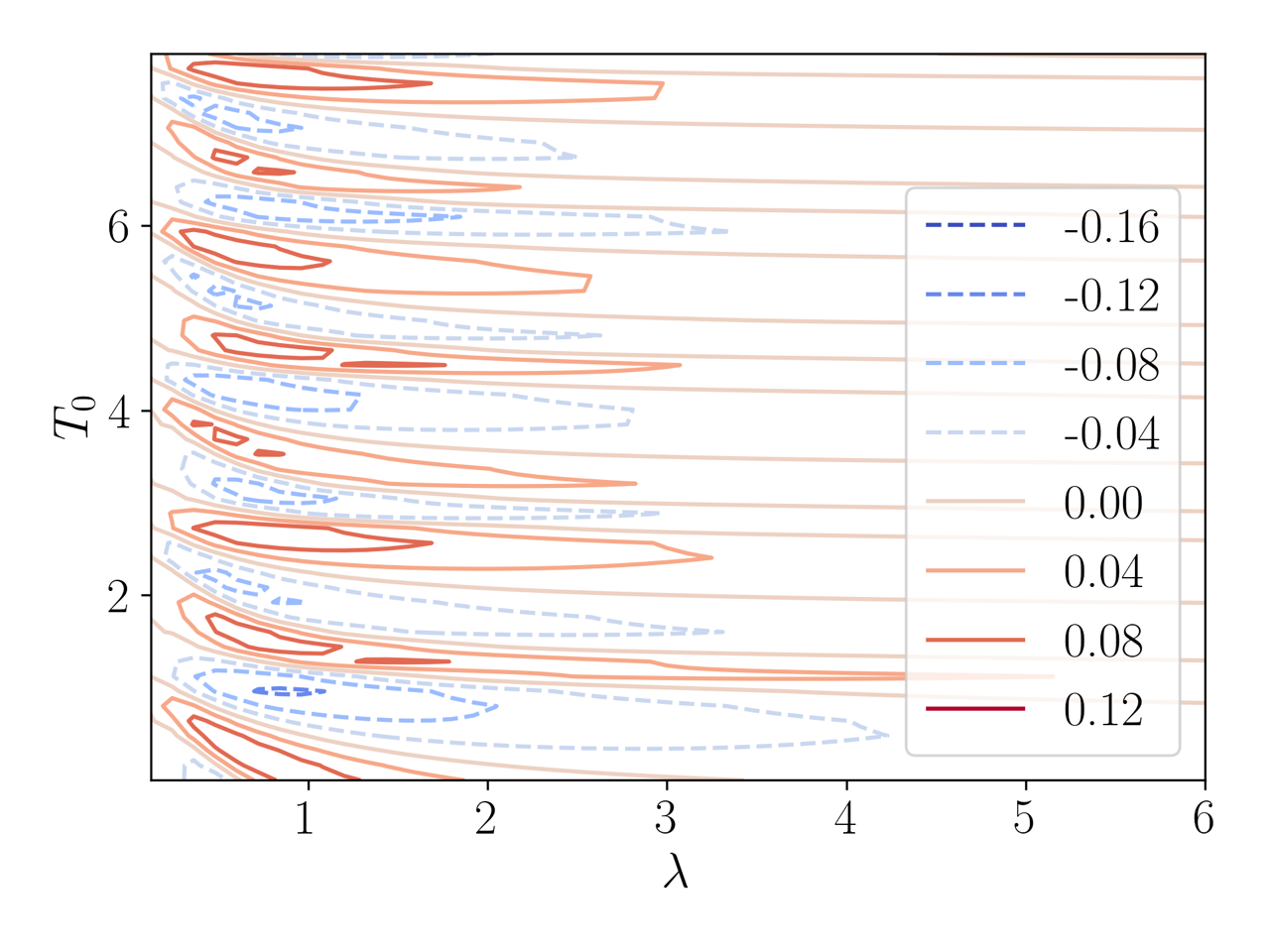}
\caption{Contour plot of $\delta\psi/\epsilon^2$ against $\tau$ and
  $x_\lambda$ from the best subcritical $\epsilon^2=0.5$ evolution of
  Baumgarte \cite{Baumgarte2018}. This is the $\epsilon^2=0.5$
  equivalent of Fig.~11 of \cite{Baumgarte2018}. $(\tau,x_\lambda)$
  are called $(T_0,\lambda)$ in this plot. We have chosen a contour
  plot rather than a surface plot as this is clearer, given the coarse
  resolution of the available data.}
\label{fig:eps05_psi2_lambda_Baumgarte}
\end{figure}

\begin{figure}
\includegraphics[scale=0.8, angle=0]{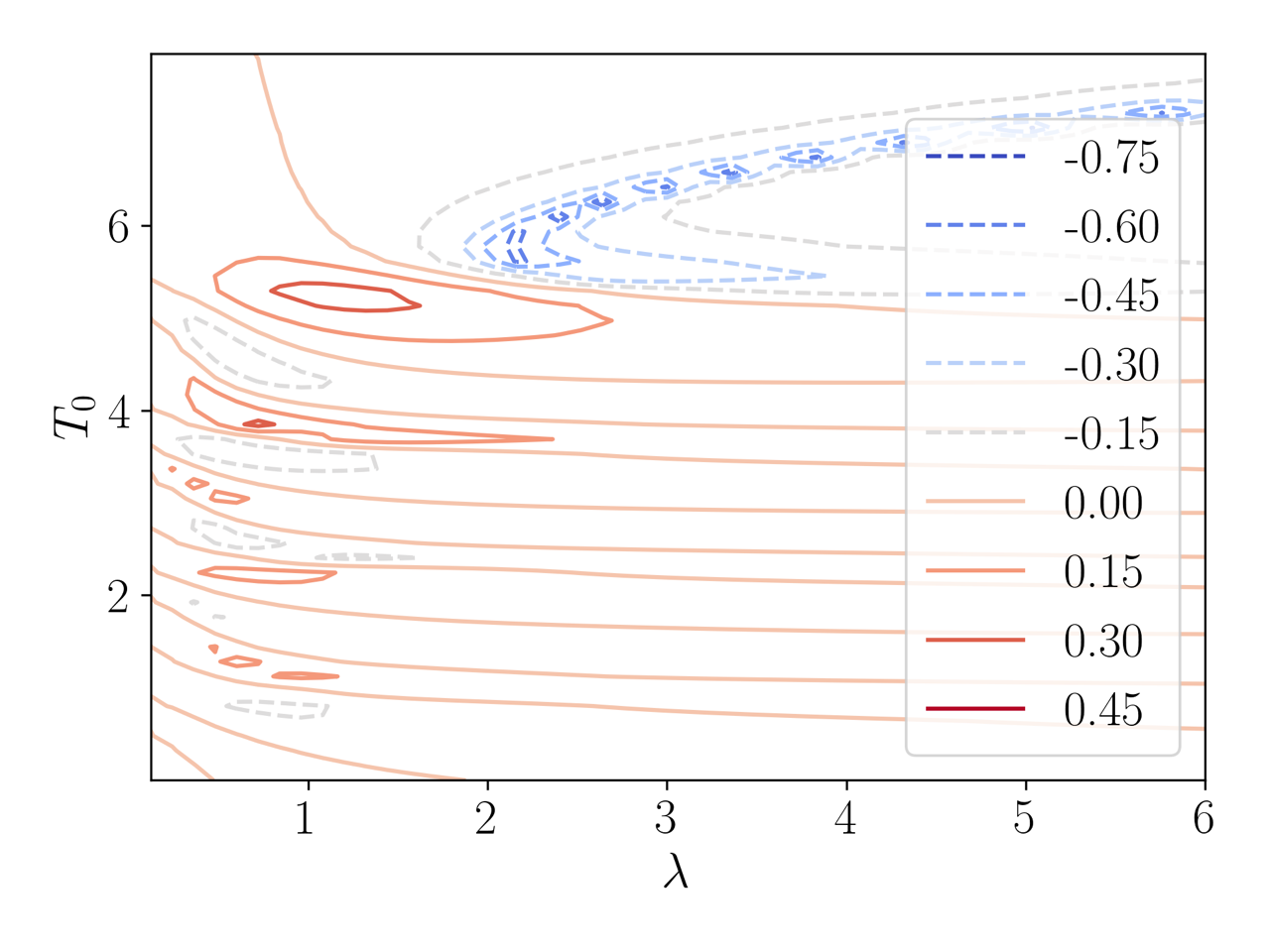}
\caption{As in Fig.~\ref{fig:eps05_psi2_lambda_Baumgarte}, but now for
  $\epsilon^2=0.75$. Note the
  sharp timelike feature in the range $6\le T_0\le 8$ that signals the
  end of approximate DSS (and, we expect, the formation of two centres
  of collapse).}
\label{fig:eps075_psi2_lambda_Baumgarte}
\end{figure}

Fig.~\ref{fig:eps075_psi2_lambda} shows $\delta\psi$, approximated as
$(3/2)\psi_2$ for our $\epsilon_2=0.75$. For comparison,
Figs.~\ref{fig:eps05_psi2_lambda_Baumgarte} and
\ref{fig:eps075_psi2_lambda_Baumgarte} show plots of $\delta\psi$
created from the simulations of \cite{Baumgarte2018} for
$\epsilon^2=0.5$ and $0.75$, respectively. This suggests that, during
the approximately self-similar phase, our $\epsilon_2=0.75$ family is
about as non-spherical as $\epsilon^2=0.5$ of \cite{Baumgarte2018}. In
both families $\delta\psi$ does not grow noticeably. By contrast,
Fig.~\ref{fig:eps075_psi2_lambda_Baumgarte} shows clear growth of
$\delta\psi$ with $\tau$ for $\epsilon^2=0.75$ and an end to the
perturbative regime around $\tau=6$.

Table~\ref{table:allcomparison} compares the black hole threshold in
selected families taken from \cite{Choptuiketal2003},
\cite{Baumgarte2018}, \cite{Maroudaetal2024} and the present
paper. Families that should be directly comparable are grouped
together. For comparison have translated $\delta\psi/\epsilon^2$ and
$\psi_2$ into $\delta\psi$ (by multiplying by $\epsilon^2$ and $3/2$,
respectively). The values for $\max|\delta\psi|$ from the null code
and that of \cite{Baumgarte2018} are read off from plots shown in the
present paper. The values for $\epsilon^2=0$ and $\epsilon^2=2/3$ of
\cite{Choptuiketal2003} are read off from Figs.~6 and 7 of that paper,
respectively. In the spherically symmetric case we include
$\max|\delta\psi|$ as an estimate of the numerical error generated by
evolving a spherically symmetric solution in cylindrical coordinates.
Finally, the values for families II, III and IV of
\cite{Maroudaetal2024} are read off from plots communicated by the
authors.

A range given for $\max|\delta\psi|$ means that its value increases
noticeably from the beginning to the end of the phase where the
spherical part of the solution is approximated by the Choptuik
solution. A single value implies that the amplitude does not grow
noticeably. (This is estimated by eye, as the perturbations are only
quasiperiodic.) 

By contrast, the ranges given for $\delta\gamma$ and $\delta\Delta$
express uncertainty: in the four papers, up to three different methods
have been used to estimate $\Delta$ (but only one for $\gamma$).
Independently, one can use either the numerical values of $\gamma$ and
$\Delta$ obtained in spherical symmetry in each paper, or their known
exact values as reference values. We cite here the largest and
smallest of these (up to six) possible differences, in order to 
indicate a plausible interval for $\delta\gamma$ and
$\delta\Delta$ for each family. We do not include the fitting errors
given in \cite{Choptuiketal2003,Maroudaetal2024}, as these appear to
be smaller than the systematic errors suggested by the intervals just
mentioned. 

We have attempted to order the families across papers, taking as our
first ordering criterion the bifurcation (yes or no) of the critical
solution, as our second criterion $\max|\delta\psi|$ (where we have
data), and otherwise $\delta\gamma$ and $\delta\Delta$.

The fact that we are able to order the families of initial data
consistently (within the estimated plausible ranges) is one of the
main physics results of this paper. It supports the
hypothesis that all near-critical evolutions go through a phase where
the solution can be approximated by the Choptuik solution plus the
growing spherical mode and the least damped $l=2$ mode. Large
nonsphericity seems to change the values of $\Delta$ and $\gamma$, as
well as the decay rate $\kappa$ of the $l=2$ mode, such that for
non-sphericities of $\max|\delta\psi|\gtrsim 0.1$ it actually
grows. (Compare this with $\max|\psi_*|\simeq 0.6$ for
the critical solution.)

The exception from this consistent picture are the values of
$\delta\gamma$ and $\delta\Delta$ obtained in this paper. Up to our
$\epsilon_2=0.5$ family, these are zero within our estimated numerical
errors. For our $\epsilon_2=0.75$ family, we have a signifant but
small change of $\delta\Delta\sim 10^{-3}$ (but not of
$\delta\gamma$), but this is still an order of magnitude smaller than
$\delta\Delta$ observed in the other three papers at what we take to
be comparable non-sphericity. For lack of a better explanation we
suspect that this is due to some systematic numerical error, more
likely in the null code than the other three codes.  

\begin{table*}
\setlength{\tabcolsep}{9pt} 
\renewcommand{\arraystretch}{1.5} 
\begin{tabular}{c|c|c|c|c|c}
Paper & Family & $\max|\delta\psi|$ & $\delta\gamma$ & $\delta\Delta$
& Bif. \\
\hline
Choptuik+ & $\epsilon^2=0$ & $0_{(i)}...0.05_{(f)}$ &
$0.008_{(e)}$ & $0_{(1,e)}...0.05_{(2,e)}$ & no \\
Baumgarte & $\epsilon^2=0$ & 0 & $0_{(e)}$ & $0.02_{(3,e)}...0.03_{(1,e)}$ & no \\
Marouda+ & I & --- & $-0.004_{(e)}$ & $-0.02_{(3,e)}...0.01_{(1,e)}$ & no \\
{\bf this paper} & $\epsilon_2=0$ & $0$ & $0_{(e)}$ & $0_{(e)}$ & no \\
\hline
Baumgarte & $\epsilon^2=10^{-2}$ & $0.005$ & $0_{(e)}$ & $-0.02_{(3,n12)}...0.03_{(1,e)}$ & no \\
\hline
{\bf this paper} & $\epsilon_2=10^{-2}$ & $0.008$ & $<10^{-3}$ & $<10^{-3}$ & no \\
\hline
Marouda+ & II & $0.02$ & $-0.002_{(e)}...0.002_{(n)}$ & $-0.07_{(3,e)}...0.03_{(2,e)}$ & no\\
\hline
Marouda+ & III & $0.06$ & $-0.010_{(e)}...-0.006_{(n)}$ & $-0.12_{(3,e)}...-0.05_{(2,e)}$ & no\\
\hline
{\bf this paper} & $\epsilon_2=0.5$ & $0.07$ & $<10^{-3}$ &
$<10^{-3}$ & no \\
\hline
Choptuik+ & $\epsilon^2=0.5$ & --- & $-0.007_{(n)}...0.001_{(e)}$ &
$-0.12_{(1,n2)}...-0.05_{(2,e)}$ & no \\
Baumgarte & $\epsilon^2=0.5$ & $0.08$ & $-0.005_{(e)}$ & 
$-0.12_{(3,n1)}...-0.05_{(1,e)}$  &  no \\
\hline
Choptuik+ & $\epsilon^2=2/3$ & $0.06_{(i)}...0.12_{(f)}$ &
 $-0.036_{(n)}...-0.028_{(e)}$ &  $-0.41_{(2,n2)}-0.31_{(1,e)}$ & no \\
\hline
{\bf this paper} & $\epsilon_2=0.75$ & $0.09$ & $-0.0016_{(e)}$ & $-0.023$ & no \\
\hline
Marouda+ & IV & $0.1_{(i)}...0.2_{(f)}$ &
$-0.045_{(e)}...-0.041_{(n)}$ & $-0.49_{(1,e)}...-0.35_{(2,e)}$ & yes
\\
\hline
Choptuik+ & $\epsilon^2=0.75$ & --- & $-0.069_{(e)}...-0.061_{(n)}$ & $-0.62_{(1,n2)}...-0.41_{(2,e)}$  &
yes \\
Baumgarte & $\epsilon^2=0.75$ & $0.11_{(i)}...0.34_{(f)}$ &
$-0.068_{(e)}$ &  $-0.72_{(3,n1)}...-0.57_{(1,e)}$ & yes \\
\hline
Choptuik+ & $\epsilon^2=5/6$ & --- & $-0.102_{(n)}...-0.094_{(e)}$ &
$-2.49_{(2,n2)}...-1.44_{(1,e)}$
& yes \\
\hline
\end{tabular}
\caption{Comparison of families of initial data near the black hole
  threshold, combining data from our evolutions with those of
  \cite{Choptuiketal2003}, \cite{Baumgarte2018} and
  \cite{Maroudaetal2024}. Families corresponding to the same initial
  data have been grouped together. A dash means no data are
  available. A range given for $\max|\delta\psi|$ means that its value
  increases noticeably during the phase where the spherical part of
  the solution is approximated by the Choptuik solution. By contrast,
  the ranges given for $\delta\gamma$ and $\delta\Delta$ express
  uncertainty, see the main text for details. Footnotes: (e) relative
  to exact values $\Delta\simeq 0.374$ and $\Delta\simeq 3.44$ in
  spherical symmetry; (n) relative to numerical value obtained in
  spherical symmetry; (i) initial and (f) final value in the
  approximately Choptuik phase; (1) first, (2) second and (3) third
  method of determining $\Delta$; (3,n1), for example means, the
  numerical value of $\Delta$ in a non-spherical evolution obtained by
  the third method, relative to the numerical value determined by the
  first method in spherical symmetry.}
\label{table:allcomparison}
\end{table*}

\subsection{Numerical problems at larger non-sphericity}


We have already mentioned that to complete the bisection in $p$ for
$\epsilon_2=0.75$ at our highest resolution $\bar N_y=17$, $\Delta
x=0.025$, we had to lower our heuristic diagnostic of collapse to $C\ge
0.8$. 

We have tried to bisect at $\epsilon_2=0.79$, with $\bar N_y=17$,
$\Delta x=0.05$, and again using $C\ge 0.8$ as the collapse
diagnostic, starting again with the initial bracket
$0.2<p_*<0.3$. Already at the third bisection step, the code stops
because $B$ becomes very large at the boundary, and so the time step
becomes very small. On closer inspection, $B$ becomes very irregular
at late times at the outermost few points, and very much larger at the
outermost grid point than anywhere else. This cannot be a boundary
instability in the usual sense because the outer boundary is treated
exactly like an ordinary grid point. Hence the instability must be one
that grows much faster at larger $x$, and so most rapidly at the outer
boundary.

We believe what is at fault here is a combination of two problems we
have already mentioned. One is that in our gauge, we evaluate
$\min_y\Xi R(u,x,y)$, and this means that the $B_{,x}$ becomes
discontinuous. This lack of smoothness then propagates to the other
variables. A separate problem is that $R_{,x}$ becomes small as
outgoing light cones are trying to recollapse, while lsB gauge (or any
gauge which solves the Raychaudhuri equation along null generators for
$G$ rather than for $R$) requires $R_{,x}>0$. We believe this would
still give rise to very large values of $B$ even if we found a better
gauge within the lsB family that avoided the unsmoothness.


\section{Conclusions}
\label{section:conclusions}


This work was born out of an attempt to generalise the method of
Garfinkle \cite{Garfinkle1995} for simulating critical collapse
without the need for adaptive mesh refinement beyond spherical
symmetry. We have demonstrated that this is possible, in the example
of axisymmetric scalar field critical collapse.

We have attempted to duplicate previous physics results as well.
Recall that the collapse simulations of Baumgarte (in spherical polar
coordinates) \cite{Baumgarte2018} seemed to have reconciled the linear
perturbation results of Mart\'\i n-Garc\'\i a and Gundlach
\cite{JMMGundlach1999} with the collapse simulations of Choptuik {\it
  et al} \cite{Choptuiketal2003} (in cylindrical coordinates): small
deviations from spherical symmetry decay when one fine-tunes to the
threshold of collapse, while large perturbations grow and lead to the
formation of two centres of collapse. This conclusion is also
supported by the more recent evolutions of \cite{Maroudaetal2024} and
(less quantitatively) of \cite{Reid2023}.

Specifically, \cite{Baumgarte2018} and \cite{Choptuiketal2003} find
that non-sphericities decay for values of their non-sphericity
parameter $\epsilon^2\le 0.5$, and \cite{Choptuiketal2003} still find
this at $\epsilon^2=2/3$. Both \cite{Baumgarte2018} and
\cite{Choptuiketal2003} find a bifurcation for $\epsilon^2=0.75$, and
\cite{Choptuiketal2003} also at $\epsilon^2=5/6$. (We note that
\cite{Maroudaetal2024} also observe bifurcation in their most
non-spherical data for the complex scalar field).
\cite{Choptuiketal2003}, \cite{Baumgarte2018} and
\cite{Maroudaetal2024} also found that $\Delta$ and $\gamma$ (measured
before any bifurcation occurs) decrease with increasing $\epsilon_2$, see
Table~1 for representative numerical values.

By contrast, we have found that even in the evolution of our
$\epsilon_2=0.75$ family of data perturbation theory is a good model
for the non-sphericity. In particular, the echoing period and critical
exponent remain at $\Delta\simeq 3.44$ and $\gamma\simeq 0.374$, their
values in spherical symmetry, and the $l=2$ non-spherical components
of all fields decay at roughly the rate predicted for linear
perturbations in \cite{JMMGundlach1999}. We also find that the
amplitude of the $l=2$ field components is almost linear in
$\epsilon_2$ for values between $10^{-4}$ and $0.75$, enhanced only by
a factor $\sim 1.5$ at $\epsilon_2=0.75$. Similarly, the $l=4$
non-sphericity scales with $(\epsilon_2)^2$, as one would expect from
perturbation theory to quadratic order, given that $l=4$ is absent in
the initial data to linear order in $\epsilon_2$, and that
\cite{JMMGundlach1999} predicts $l=4$ to decay much more rapidly than
$l=2$.

However, we cannot evolve the same family of initial data as
\cite{Baumgarte2018} and \cite{Choptuiketal2003}, because we set data
on an outgoing null cone. We have compared a gauge-invariant measure
of the non-sphericity of the scalar field, in the phase where a
near-critical evolution is approximately self-similar, with the same
measure for the evolutions of \cite{Baumgarte2018}, and find that this
comparable between our $\epsilon_2=0.5$ and $\epsilon^2=0.75$ of
\cite{Baumgarte2018} (and by implication of
\cite{Choptuiketal2003}). If we compare our $\epsilon_2=0.5$ with
$\epsilon^2=0.75$ of \cite{Baumgarte2018,Choptuiketal2003}, a mild
tension remains, as these authors claim that $\Delta\simeq 3.44$ is
reduced by $\sim 0.08$ and $\gamma\simeq 0.374$ by $\sim 0.007$,
whereas we see much smaller reductions. This could potentially still
be explained by numerical error.

In order to accurately measure the critical exponent $\gamma$, we have
fitted not only the power laws with exponent $\gamma$, but also the
small periodic fine-structure superimposed on them. This has only been
done in spherical symmetry before. Our fine-structure of the curvature
scaling laws (on the subcritical side) agrees well with published
results in spherical symmetry, but our fine structure of the mass
scaling law does not. The reason may be that our collapse criterion is
not the first appearance of a trapped surface (in a given time
slicing) but the first appearance of a coordinate 2-surface ${\cal
  S}_{u,x}$ with Hawking compactness above a threshold value, and we
evaluate the Hawking mass of that surface.

We believe this is the first simulation of gravitational collapse
beyond spherical symmetry in null coordinates. (The paper
\cite{Siebeletal2003} investigates axisymmetric supernova core
collapse, but not black hole formation.) We have applied our methods
to the challenging problem of axisymmetric scalar field critical
collapse, and at moderate non-sphericity have been able to fine-tune
our initial data to the black-hole threshold to machine precision,
without the need for mesh refinement.  

We have not yet been able to simulate critical collapse for
large enough non-sphericity to fully duplicate the results of
\cite{Baumgarte2018,Choptuiketal2003,Maroudaetal2024}: at large
non-sphericity, our evolutions
stop before we can classify them as either forming a black hole or
dispersing. The proximate cause for this is that the divergence of the null
generators of our coordinate lightcones is trying to become negative
at some points on our last null cone, but the generalised Bondi
coordinates we use break down if this happens. Therefore, the next
step will be to change to a generalised affine radial coordinate. This
will then allow us to investigate if our null cones themselves remain
regular in critical collapse, or form caustics. 


\acknowledgments

We would like to thank Bernd Br\"ugmann, Tom\'a\v{s} Ledvinka, Anton
Khirnov, Daniela Cors and Krinio Marouda for helpful discussions, and
the Mathematical Research Institute Oberwolfach for supporting this
work through its ``Oberwolfach Research Fellows'' scheme.  DH was
supported in part by FCT (Portugal) Project No. UIDB/00099/2020. TWB
was supported in part by National Science Foundation (NSF) grants
PHY-2010394 and PHY-2341984 to Bowdoin College.


\appendix



\section{Family-dependent parameters in the fine structure of the scaling laws}
\label{appendix:wiggles}


In this Appendix, we show that the parameters $A$ and $B$ in
Eqs.~(\ref{curvscaling}) and (\ref{massscaling}) are family-dependent, and
independent of each other.

Focus first on the late-time evolution of the exactly-critical member
of a given 1-parameter family of initial data. If we fix a
(family-independent) small length scale $L$, then we can, for example,
measure the phase of the scalar field $\phi$ at the centre when the
Ricci scalar at the centre takes exactly the value $L^{-2}$. This is
equivalent to our $A$.

On the other hand, to derive the scaling laws we use the fact that in
near-critical evolutions the amplitude of the growing perturbation is,
to leading order, proportional to $(p-p_*)e^{\lambda_0\tau}$. The constant of
proportionality must again be family-dependent, already for the
trivial reason that $p$ can have any dimension. Its logarithm is our $B$.

Note that for perfectly-critical initial data the growing perturbation
of the critical solution is by construction absent, while the scaling
laws rely on the amplitude of the growing mode. Hence the
family-dependent constants $A$ and $B$ are independent of each
other. 


\section{Convergence tests in the strong subcritical regime}
\label{appendix:convergencetestsold}


For convergence testing we choose our $\epsilon_2=0.75$ initial data
with the two amplitudes that we have also used as an initial bracket
of the black hole threshold, namely $p=0.2$ (subcritical) and $p=0.3$
(supercritical). We use the grid parameters $x_\text{max}=50$ and
$x_0=33$ throughout, which we also used in our near-critical evolutions.

For $p=0.2$ gravity is strong but not close to collapse, with a
maximum Hawking compactness of $\bar C\simeq 0.08$ in the initial
data, and reaching $\bar C\simeq 0.2$ during the evolution (compared
to $\bar C\simeq 0.6$ in the critical solution and $\bar C=1$ on a
black-hole horizon). We run to $u=6$, when the solution has largely dispersed,
with $\bar C<0.01$, but the numerical domain has 
contracted from $R_\text{max}=25$ at the initial time only to
$R_\text{max}\simeq 14$, thus avoiding very small timesteps. 

The evolutions with $p=0.3$ start from $\bar C\simeq 0.18$ and reach
our heuristic collapse criterion $\bar C=0.8$ at $u\simeq 3.9$, at
which point we stop the evolution.

Both $p=0.2$ and $p=0.3$ are far enough from $p_*\simeq 0.25$ that it
remains meaningful to compare numerical solutions with different
resolutions at the same coordinate time $u$ and amplitude $p$, as
$|p_*-p|\simeq 0.05$ is then much larger than the variation of $p_*$
with different numerical parameters. By contrast, in the critical
regime we would need to compare different resolutions at the same
(small) $p/p_*-1$, with $p_*$ resolution-dependent, and plot them
against the similarity-adapted coordinates $(\tau,\xi)$, which also depend
sensitively on numerical parameters through the fitted value of $u_*$.

We output all fields as $l$-components (rather than against $y$) to
make comparisons at different $\bar N_y$ and $l_\text{max}$ simpler.
We have tested for self-convergence to second order in $\Delta x$ and
in $l_\text{max}$, see Paper~I for details of how the errors presented
here are defined.

We have evolved at two intersecting families of resolutions: at
$\Delta x=\{0.1,{\bf 0.05},0.025,0.0125\}$ with fixed angular
resolution $\bar N_y=17$, $l_\text{max}=16$; and at
$\bar N_y=\{9,13,{\bf 17},25,33,49,65\}$, $l_\text{max}=\bar N_y-1$,
with fixed radial resolution $\Delta x=0.05$. Their intersection is
our baseline solution, chosen because the discretisation errors in $x$
and $y$ are roughly similar. Note that our adaptive timestep
$\Delta u$ is approximately proportional to $\Delta x$ and
approximately independent of angular resolution.

We find a large and non-converging error at small $u$ and $x$, in all
$\psi_l$ with $l>2$. We believe this is due to the fact that our
initial data (\ref{Phieps}-\ref{psieps}) is not actually single-valued
at the origin. The non-convergent error disappears quickly, and we
believe this is because the boundary conditions at the centre force
the solution to become single-valued.

\begin{figure}
\includegraphics[scale=0.67, angle=0]{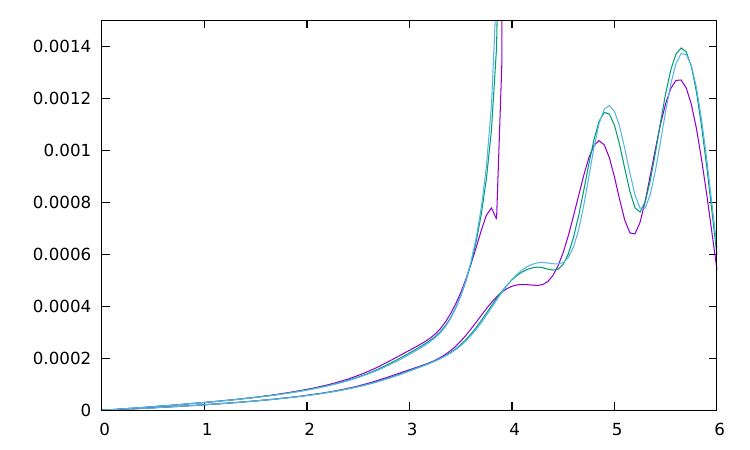} 
\includegraphics[scale=0.67, angle=0]{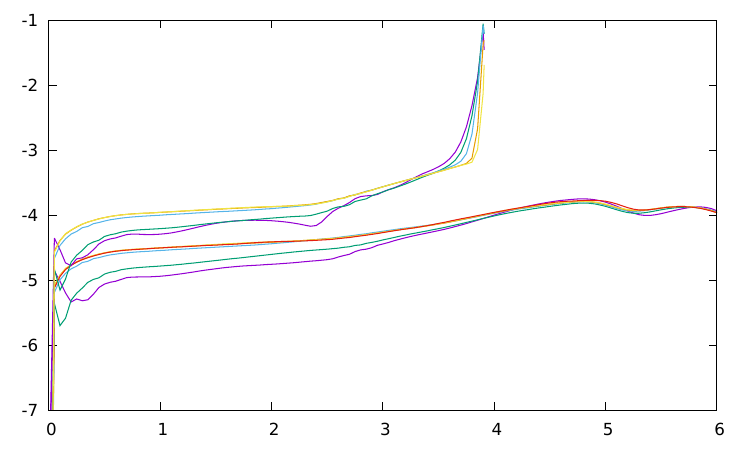} 
\caption{{\em Upper plot}: $L^2$ norm of the discretisation error in
  $x$ of the spherical scalar field component $\psi_0(u,x)$, shown
  against $u$, for $\Delta x=\{0.1,0.05,0.25\}$ (purple, green, blue),
  scaled to the baseline resolution $\Delta x=0.05$ by assuming that
  the error scales as $\Delta x^2$. The angular resolution is fixed at
  $\bar N_y=17$, $l_\text{max}=16$.  The lower group of three curves
  is for the $p=0.2$ evolutions, and the upper group for the $p=0.3$
  evolutions.  {\em Lower plot}: $\log_{10}$ of the $L^2$ norm of the
  discretisation error in $y$ for $\bar N_y=\{9,13, 17,25,33,49\}$
  (purple, green, blue, orange, yellow, red) scaled to the baseline
  resolution $\bar N_y=17$, $l_\text{max}=16$ assuming that the error
  scales as $l_\text{max}^{-2}$. The radial resolution is fixed at
  $\Delta x=0.05$.}
\label{fig:eps075_weakstrong_conv_psi00}
\end{figure}

\begin{figure}
\includegraphics[scale=0.67, angle=0]{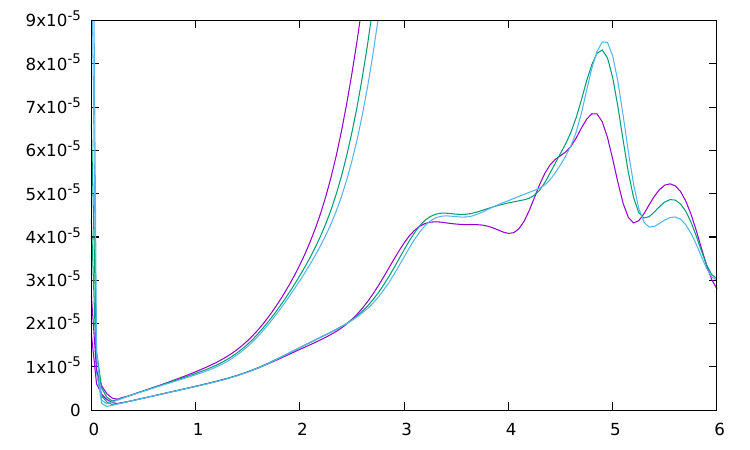}
\includegraphics[scale=0.67, angle=0]{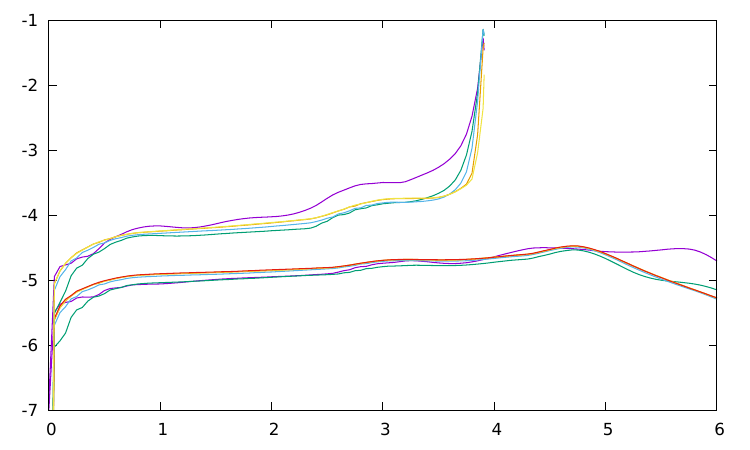}
\caption{As in Fig.~\ref{fig:eps075_weakstrong_conv_psi00}, but now for
  $\psi_4$. The large spike at $u=0$ (not fully shown, and not
  convergent) is a signal of the irregularity of our initial
  conditions. All $\psi_l$ with $l>2$ have this problem.}
\label{fig:eps075_weakstrong_conv_psi04}
\end{figure}

\begin{figure}
\includegraphics[scale=0.67, angle=0]{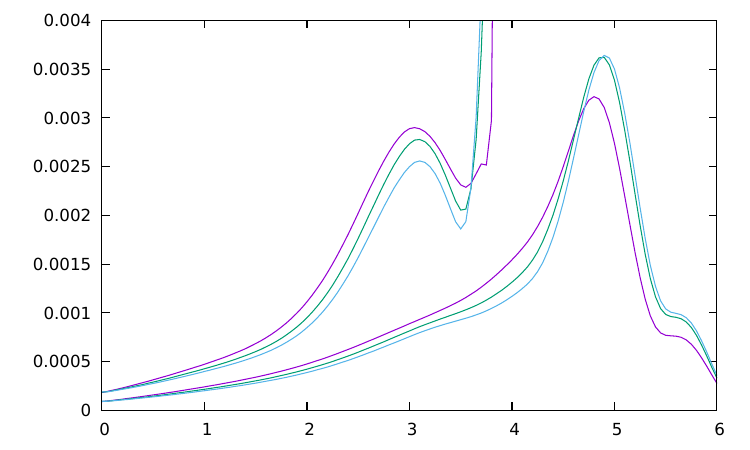}
\includegraphics[scale=0.67, angle=0]{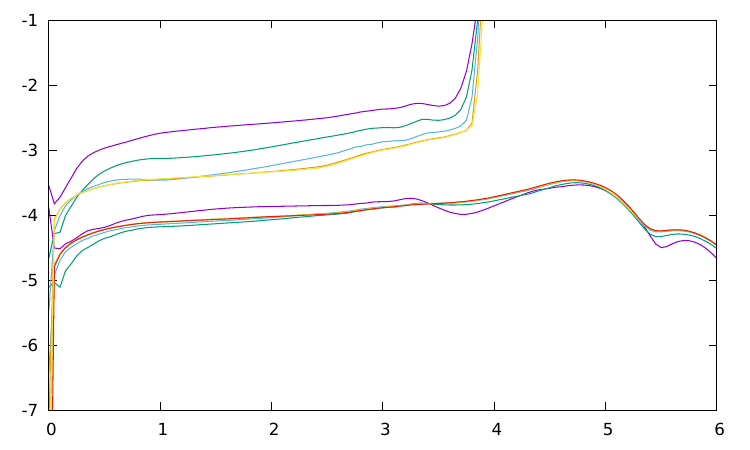}
\caption{As in Fig.~\ref{fig:eps075_weakstrong_conv_psi00}, but now for
  $\bar C$. }
\label{fig:eps075_weakstrong_conv_twoMtestoR}
\end{figure}

At all other $(u,x)$ we find the expected second-order convergence in
$\Delta x$.  For a spectral method acting on a smooth solution, we
would expect approximately exponential convergence, but in fact the
discretisation error in $y$ decreases with resolution only as
$l_\text{max}^{-2}$.  Second-order convergence in the $L^2$ norm with
$\Delta x$ and $1/l_\text{max}$ is demonstrated for $\psi_0$ in
Fig.~\ref{fig:eps075_weakstrong_conv_psi00}, for $\psi_4$ in
Fig.~\ref{fig:eps075_weakstrong_conv_psi04}, and for $\bar C$ in
Fig.~\ref{fig:eps075_weakstrong_conv_twoMtestoR}. The convergence is
actually pointwise. The global maximal errors $\max_u ||{\cal
  E}_x||_{L^2}$ and $\max_u||{\cal E}_y||_{L^2}$ for the dispersing
are shown in Table~\ref{table:errors}. Given second-order convergence,
we can accurately estimate the error at all higher resolutions by
scaling these numbers by factors $(\Delta x/0.05)^2$ and
$(l_\text{max}/16)^{-2}$, respectively.

The computation of the diagnostic $\bar C$ in the collapsing solution
seems to be particularly challenging. Here we see clear second-order
convergence, even at early times, only for $\bar N_y\ge 25$, see the
lower plot in Fig.~\ref{fig:eps075_weakstrong_conv_twoMtestoR}.

\begin{table}
\setlength{\tabcolsep}{6pt} 
\renewcommand{\arraystretch}{1.3} 
\begin{tabular}{l|lrrrr}
$\epsilon_2$ & $\Delta x$ & $\bar N_y$ & $l_\text{max}$ & $x_0$ &
  $x_\text{max}$  \\
\hline
0.5 & 0.025 & 9/17 & 8/16 & 29 & 30  \\
0.6 & 0.025 & 9/17 & 8/16 & 39 & 40  \\
0.7 & 0.05  & 17   & 16   & 99 & 100  \\
0.75 & 0.05 & 17   & 16   & 149 & 150 \\
\end{tabular}
\caption{Successful initial bracketings of the collapse threshold with
  the analytic initial data (\ref{analytictrydata}). In each case,
  $p=0.2$ disperses and $p=0.3$ collapses. $x_\text{max}$ needs to be
  this large only for the collapsing solutions. $x_\text{max}$ is
  twice as large as for the non-analytic data (\ref{trydata}) for
  $\epsilon_2=0.5$, and three times as large for
  $\epsilon_2=0.75$. $x_0$ has not yet been fine-tuned for bisection
  to the black hole threshold. We have used $\bar C\ge 0.8$ as the
  collapse threshold and $\bar C\le 0.05$ as the dispersion
  threshold.}
\label{table:new3p10epsruns}
\end{table}

\begin{table}
\setlength{\tabcolsep}{6pt} 
\renewcommand{\arraystretch}{1.3} 
\begin{tabular}{c|ccc}
non-analytic & $\psi_0$ & $\psi_4$ & $\bar C$ \\
\hline
$\max_u ||{\cal E}_x||_{L^2}$ & $1.3\cdot 10^{-3}$ & $8 \cdot
10^{-5}$ & $3.5\cdot 10^{-3}$ \\
$\max_u||{\cal E}_y||_{L^2}$  & $1.7\cdot 10^{-4}$ & $3\cdot 10^{-5}$ & $3\cdot
10^{-4}$ \\
\hline
\hline
analytic & $\psi_0$ & $\psi_4$ & $\bar C$ \\
\hline
$\max_u ||{\cal E}_x||_{L^2}$ & $2\cdot10^{-3}$ & $5\cdot10^{-4}$ & $7\cdot 10^{-3}$ \\
$\max_u||{\cal E}_y||_{L^2}$  & $1.2\cdot10^{-2}$ & $4\cdot10^{-3}$ & $2\cdot10^{-2}$ \\
\end{tabular}
\caption{Table of the maximum in $u$ of the $L^2$ norm in $x$ and $y$ of
  the discretisation errors in $x$ and $y$, in the three
  variables $\psi_0$, $\psi_0$ and $\bar C$, in the $p=0.2$
  (dispersing) solution, at the baseline resolution $\Delta x=0.0.5$,
  $\bar N_y=17$, $l_\text{max}=16$. Top: non-analytic data. Bottom:
  analytic data.}
\label{table:errors}
\end{table}

\begin{figure}
\includegraphics[scale=0.67, angle=0]{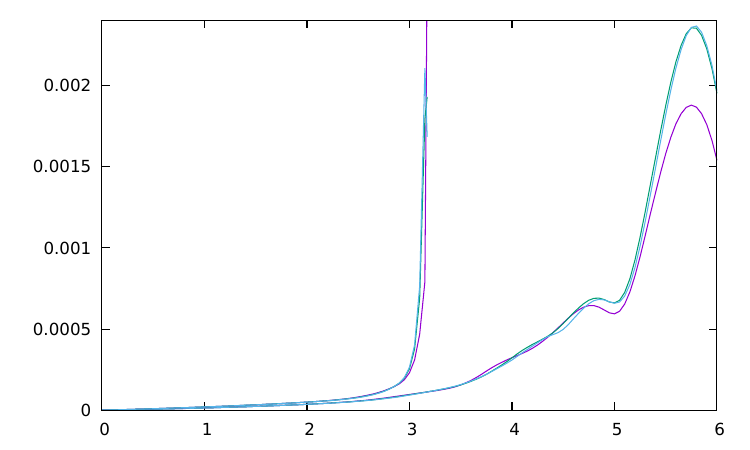} 
\includegraphics[scale=0.67, angle=0]{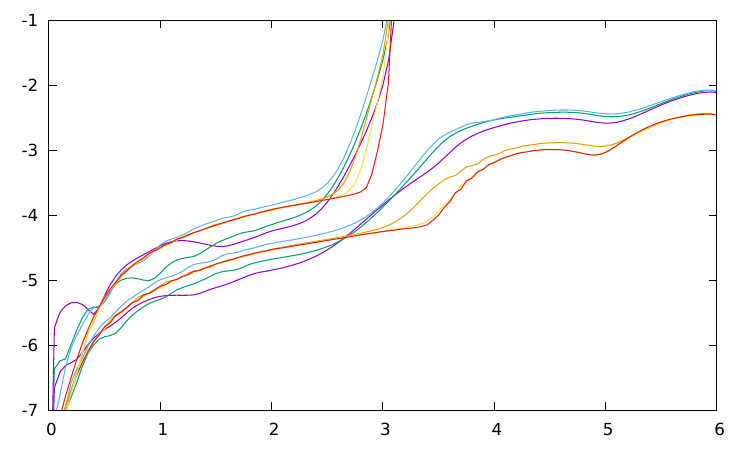} 
\caption{Error in $\psi_0$ as in
  Fig.~\ref{fig:eps075_weakstrong_conv_psi00}, but now for the
  analytic initial data (\ref{analytictrydata}) with
  $\epsilon_2=0.75$, and with
  errors in $y$ now scaled by $\exp[-\kappa(l_\text{max}-16)]$ with
  $\kappa\simeq 0.280$ for $l_\text{max}=8,12,16$ and
  $\kappa\simeq 0.085$ for $l_\text{max}=24,32,48$.}
\label{fig:new3p10eps075_weakstrong_conv_psi00}
\end{figure}

\begin{figure}
\includegraphics[scale=0.67, angle=0]{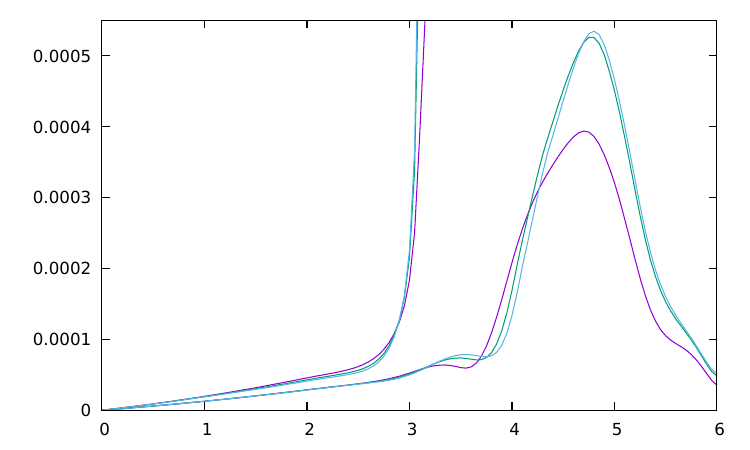}
\includegraphics[scale=0.67, angle=0]{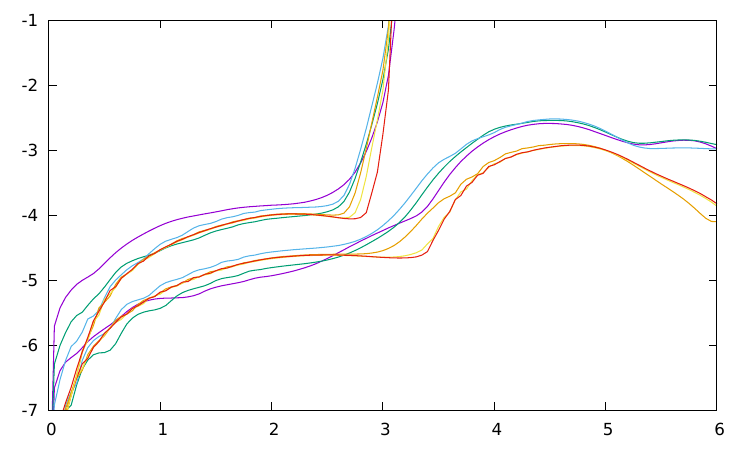}
\caption{As in Fig.~\ref{fig:new3p10eps075_weakstrong_conv_psi00}, but now for
  $\psi_4$.}
\label{fig:new3p10eps075_weakstrong_conv_psi04}
\end{figure}

\begin{figure}
\includegraphics[scale=0.67, angle=0]{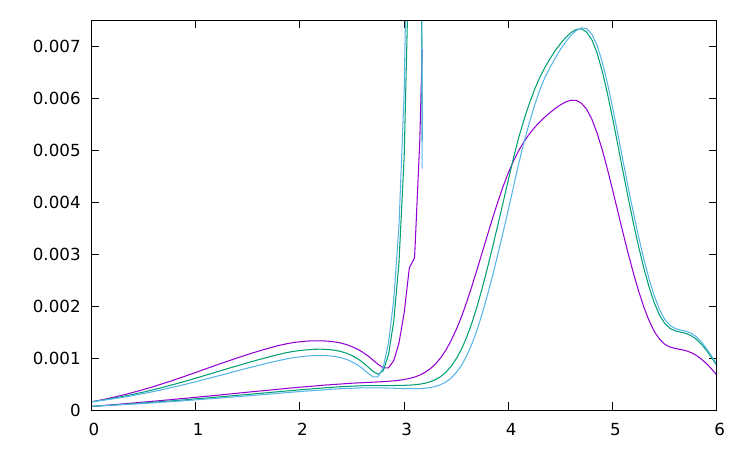}
\includegraphics[scale=0.67, angle=0]{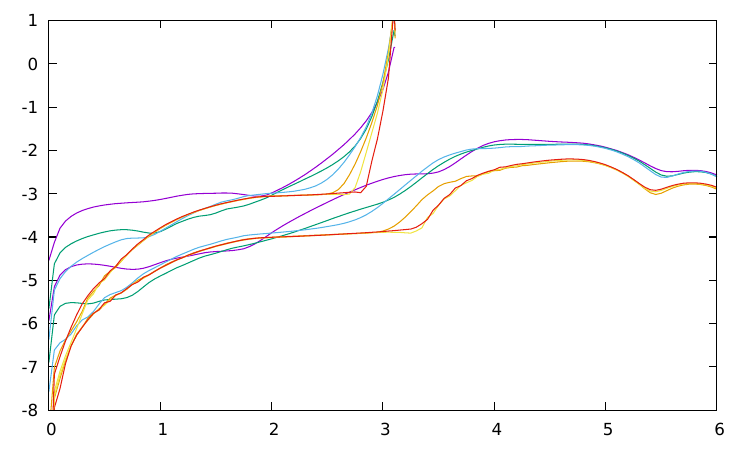}
\caption{As in Fig.~\ref{fig:new3p10eps075_weakstrong_conv_psi00}, but now for
  $\bar C$. }
\label{fig:new3p10eps075_weakstrong_conv_twoMtestoR}
\end{figure}

\begin{figure}
\includegraphics[scale=0.67, angle=0]{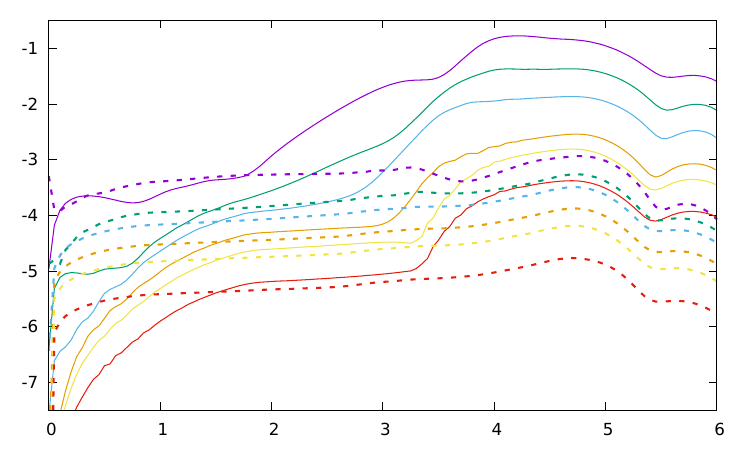}
\caption{Direct comparison of the unscaled discretization errors in
  $y$ for the non-analytic (dashed lines) and analytic (solid lines)
  $\epsilon_2=0.75$, $p=0.2$ data. As before $\bar N_y=\{9,13, 17,25,33,49\}$
  are shown in purple, green, blue, orange, yellow, red, and we
  estimate the error by subtracting the $\bar N_y=65$ evolutions.}
\label{fig:new3p10eps075_eps075_weak_convy_twoMtestoR}
\end{figure}


\section{Convergence tests with initial data that are regular at the origin}
\label{appendix:convergencetestsnew}


To check the effect of our non-analytic original choice of initial
data (where $\psi$ is not single-valued at the origin) on convergence,
we have created a second family of initial data that are analytic.

We computed the exact solution $\phi_{(2k)}(t,r,y)$ of the flat
spacetime scalar wave equation with Cauchy data
\begin{eqnarray}
\label{phipdef}
\phi_{(2k)}(0,r,y)&=&e^{-r^2}(-ry)^{2k}, \\
\phi_{(2k),t}(0,r,y)&=&0,
\end{eqnarray}
for $k=0,1,\dots 5$. (Recall that $y:=-\cos\theta$. We have set the
width of the Gaussian to $d=1$ to fix an overall scale). These can be
found as linear combinations of the generalised d'Alembert solutions
$\phi_l(t,r)P_l(y)$ constructed in Paper~I, for even $l$ up to $2k$,
each with an ansatz for the free function $\chi_l(r)$ of that is
$\exp(-r^2)$ times an odd polynomial in $r$ of order $l+1$.

We then use the $\phi_{(2k)}(t,r,y)$ as building blocks to construct
the solution with initial data
\begin{eqnarray}
\bar\phi(0,r,y)&=&pe^{-r^2}\sum_{k=0}^5 {(\epsilon_2r^2y^2)^k\over k!} \\
&=& 
p\,e^{-{r^2(1-\epsilon_2y^2)}} +O(\epsilon^6 r^{12} y^{12}), \\
\phi_{,t}(0,r,y)&=&0.
\end{eqnarray}
We then read off the desired null data at $u=0$ from the full
  solution as
\begin{equation}
\label{analytictrydata}
\psi(0,r,y)=\bar\phi(r,r,y).
\end{equation}

We were able to compute up to $k=6$ in Mathematica, and the error in
the Taylor series is then below machine precision. However, in
contrast to the $k\le 5$ terms, the $k=6$ term would need to be
approximated near $r=0$ to avoid large roundoff error when evaluating
the exact expression numerically in the code, so we settle for the
above truncation at $k=5$.

Table~\ref{table:new3p10epsruns} shows successful initial brackets of
the black hole threshold with the analytic data. We have not bisected
the analytic data to the black hole threshold.

For convergence testing, we have evolved $\epsilon_2=0.75$, $p=0.2$
and $p=0.3$, at the same two intersecting families of resolution as
for our non-analytic data. We again see second-order convergence in $\Delta
x$. Power-law convergence in $l_\text{max}$ is no longer a good fit. A
better fit is exponentional convergence in $l_\text{max}$ with a break
in the exponent, namely $\exp[-\kappa(l_\text{max}-16)]$ with
$\kappa\simeq 0.280$ for $l_\text{max}=8,12,16$ and $\kappa\simeq
0.085$ for $l_\text{max}=24,32,48$. 

Numerical values for the errors at the baseline resolution are shown
in Table~\ref{table:errors}, and $L^2$ norms over $x$ of the errors in
$\psi_0$, $\psi_4$ and $\bar C$ are shown in
Figs.~\ref{fig:new3p10eps075_weakstrong_conv_psi00},
\ref{fig:new3p10eps075_weakstrong_conv_psi04} and
\ref{fig:new3p10eps075_weakstrong_conv_twoMtestoR}.  For completeness,
we show the unscaled errors in $y$, for the variable $\bar C$, and for
both the analytic and non-analytic $\epsilon_2=0.75$, $p=0.2$ data in
Fig.~\ref{fig:new3p10eps075_eps075_weak_convy_twoMtestoR}.




\begin{thebibliography}{99}



\bibitem{paper1} C. Gundlach, D. Hilditch and T. W. Baumgarte,
  Simulations of gravitational collapse in null coordinates I:
  Formulation and weak-field tests in generalised Bondi gauges,
  \href{https://arxiv.org/abs/2404.15105}{arXiv:2404.15105}.

\bibitem{GundlachLRR} C. Gundlach and J. M. Mart\'\i n-Garc\'\i a,
  Critical phenomena in gravitational collapse,
  \href{https://link.springer.com/article/10.12942/lrr-2007-5}{Living
    Rev. Relativ. {\bf 10}, 5 (2007)}.

\bibitem{JMMGundlach1999} J. M. Mart\'\i n-Garc\'\i a and C. Gundlach,
  All nonspherical perturbations of the Choptuik spacetime decay, 
  \href{https://journals.aps.org/prd/abstract/10.1103/PhysRevD.59.064031}{Phys. Rev. D {\bf 59}, 064031
  (1999)}. 

\bibitem{Choptuiketal2003} M. W. Choptuik, E. W. Hirschmann,
  S. L. Liebling and F. W. Pretorius, Critical collapse of the
  massless scalar field in axisymmetry, \href{https://journals.aps.org/prd/abstract/10.1103/PhysRevD.68.044007}{Phys. Rev. D {\bf 68}, 044077
  (2003)}. 

\bibitem{Baumgarte2018} T. W. Baumgarte, Aspherical deformations of
  the Choptuik spacetime, \href{https://journals.aps.org/prd/abstract/10.1103/PhysRevD.98.084012}{Phys. Rev. D {\bf 98}, 084012 (2018)}. 

\bibitem{Choptuik1993} M. W. Choptuik, Universality and scaling in
  gravitational collapse of a massless scalar field,
  \href{https://journals.aps.org/prl/abstract/10.1103/PhysRevLett.70.9}{Phys. Rev. Lett. {\bf
      70}, 9 (1993)}.

\bibitem{MartinGundlach2003} J. M. Mart\'\i n-Garc\'\i a and
  C. Gundlach, Global structure of Choptuik’s critical solution in
  scalar field collapse,
  \href{https://journals.aps.org/prd/abstract/10.1103/PhysRevD.68.024011}{Phys. Rev. D
    {\bf 68}, 024011 (2003)}.

\bibitem{Healeyetal2014} J. Healy and P. Laguna, Critical collapse of
  scalar fields beyond axisymmetry, \href{https://link.springer.com/article/10.1007/s10714-014-1722-2}{Gen. Relativ. Gravit. {\bf 46}, 1722
    (2014)}.

\bibitem{Deppetelal2019} N. Deppe, L. E. Kidder, M.
  A. Scheel and S. A. Teukolsky, Critical behavior in 3D
  gravitational collapse of massless scalar fields, \href{https://journals.aps.org/prd/abstract/10.1103/PhysRevD.99.024018}{Phys. Rev. D
    {\bf 99}, 024018 (2019)}.

\bibitem{CloughLim2016} K. Clough and E. A. Lim, Critical Phenomena in
  Non-spherically Symmetric Scalar Bubble Collapse, unpublished, \href{https://arxiv.org/abs/1602.02568}{https://arxiv.org/abs/1602.02568}.

\bibitem{Reid2023} G. D. Reid and M. W. Choptuik, Universality in the
  critical collapse of the Einstein-Maxwell system,
  \href{https://journals.aps.org/prd/abstract/10.1103/PhysRevD.108.104021}{Phys. Rev. D
    {\bf 108}, 104021, (2023)}.

\bibitem{Maroudaetal2024} K. Marouda, D. Cors, H. R. Rüter,
  F. Atteneder and D. Hilditch, Twist-free axisymmetric critical
  collapse of a complex scalar field,
  \href{https://journals.aps.org/prd/abstract/10.1103/PhysRevD.109.124042}{Phys. Rev. D
    {\bf 109}, 124042 (2024)}

\bibitem{Garfinkle1995} D. Garfinkle, Choptuik scaling in null
  coordinates, \href{https://journals.aps.org/prd/abstract/10.1103/PhysRevD.51.5558}{Phys. Rev. D {\bf 51}, 5558
  (1995)}.

\bibitem{ymscalar} C. Gundlach, T. W. Baumgarte and D. Hilditch, Critical
  phenomena in gravitational collapse with two competing massless
  matter fields, \href{https://journals.aps.org/prd/abstract/10.1103/PhysRevD.100.104010}{Phys. Rev. D {\bf 100}, 104010 (2019)}.

\bibitem{Rinne2020} O. Rinne, Type II critical collapse on a single
  fixed grid: a gauge-driven ingoing boundary method, \href{https://link.springer.com/article/10.1007/s10714-020-02768-x}{Gen. Relativ. Gravit. {\bf 52}, 117
  (2020)}. 

\bibitem{PortoGundlach2022} B. Porto-Veronese and C. Gundlach,
  Critical phenomena in gravitational collapse with competing scalar
  field and gravitational waves in 4+1 dimensions,
  \href{https://journals.aps.org/prd/abstract/10.1103/PhysRevD.106.104044}{Phys. Rev. D
    {\bf 106}, 104044 (2022)}.

\bibitem{Gundlach1997} C. Gundlach, Understanding critical collapse of
  a scalar field, \href{https://journals.aps.org/prd/abstract/10.1103/PhysRevD.55.695}{Phys. Rev. D {\bf 55}, 695
  (1997)}. 

\bibitem{PuerrerHusaAichelburg} M. P\"urrer, S. Husa and
  P. C. Aichelburg, News from critical collapse: Bondi mass, tails,
  and quasinormal modes, \href{https://journals.aps.org/prd/abstract/10.1103/PhysRevD.71.104005}{Phys. Rev. D {\bf 71}, 104005
  (2005)}.
 
\bibitem{CrespoOliveiraWinicour2019} J. A. Crespo, H. P. de Oliveira and
  J. Winicour, The affine-null formulation of the gravitational
  equations: spherical case, \href{https://journals.aps.org/prd/abstract/10.1103/PhysRevD.100.104017}{Phys. Rev. D {\bf 100}, 104017 (2019)}. 

\bibitem{GarfinkleDuncan1998} D. Garfinkle and G. C. Duncan, Scaling
  of curvature in subcritical gravitational collapse, \href{https://journals.aps.org/prd/abstract/10.1103/PhysRevD.58.064024}{Phys. Rev. D
  {\bf 58}, 064024 (1998)}.

\bibitem{Corsetal2023} D. Cors, S. Renkhoff, H. R. R\"uter,
  D. Hilditch and B. Br\"ugmann, Formulation improvements for critical
  collapse simulations,
  \href{https://journals.aps.org/prd/abstract/10.1103/PhysRevD.108.124021}
{Phys. Rev. D {\bf 108}, 124021, (2023)}.

\bibitem{Siebeletal2003} F. Siebel, J. A. Font, E. M\"uller and
  P. Papadopoulos, Axisymmetric core collapse simulations using
  characteristic numerical relativity,
  \href{https://journals.aps.org/prd/abstract/10.1103/PhysRevD.67.124018}{Phys. 
Rev. D {\bf 67}, 124018 (2003)}.


\end{thebibliography}
\end{document}